\documentclass[a4paper,fleqn,usenatbib]{mnras}

\usepackage{newtxtext,newtxmath}
\usepackage[T1]{fontenc}
\usepackage{ae,aecompl}
\usepackage{graphicx}	
\usepackage{amsmath}	
\usepackage{amssymb}	
\usepackage{color}

\definecolor{orange}{rgb}{1,0.5,0}

\newcommand{\ud}{\mathrm{d}}
\newcommand{\pd}{\mathrm{\partial}}

\title[Gravitational Wave Emission]{Gravitational Wave Emission from 3D Explosion Models of Core-Collapse Supernovae with Low and Normal Explosion Energies}

\author[Jade Powell \& Bernhard  M\"uller]{
  Jade Powell$^{1}$\thanks{E-mail: jade.powell@ligo.org}
  and
Bernhard M\"uller$^{2}$
\\
$^{1}$OzGrav, Centre for Astrophysics and Supercomputing, Swinburne University of Technology, Hawthorn, VIC 3122, Australia.\\
$^{2}$Monash Centre for Astrophysics, School of Physics and Astronomy, Monash University, VIC 3800, Australia.\\
}

\date{Accepted 2019 May 6. Received 2019 April 12; in original form 2018 December 13}

\pubyear{2018}

\begin{document}
\label{firstpage}
\pagerange{\pageref{firstpage}--\pageref{lastpage}}
\maketitle

\begin{abstract}
 Understanding gravitational wave emission from core-collapse
 supernovae will be essential for their detection with current and
 future gravitational wave detectors. This requires a sample of waveforms from modern
 3D supernova simulations reaching well into the explosion phase,
 where gravitational wave emission is expected to peak. However, recent waveforms from
 3D simulations with multi-group neutrino transport do not reach far
 into the explosion phase, and some are still obtained from
 non-exploding models. We therefore calculate waveforms up to 0.9\,s
 after bounce using the neutrino hydrodynamics code
 \textsc{CoCoNuT-FMT}. We consider two models with low and normal
 explosion energy, namely explosions of an ultra-stripped progenitor
 with an initial helium star mass of $3.5\,M_{\odot}$, and of an
 $18\,M_{\odot}$ single star. Both models show gravitational wave emission from the
 excitation of surface g-modes in the proto-neutron star with
 frequencies between $\mathord{\sim}800\,\mathrm{Hz}$ and 1000\,Hz at
 peak emission. The peak amplitudes are about $6\, \mathrm{cm}$ and
 $10\, \mathrm{cm}$, respectively, which is somewhat higher than in most recent
 3D models of the pre-explosion or early explosion phase.  Using a Bayesian analysis, we determine the
 maximum detection distances for our models in simulated Advanced
 LIGO, Advanced Virgo, and Einstein Telescope design sensitivity
 noise. The more energetic $18 M_\odot$ explosion will be detectable
 to about $17.5 \,\mathrm{kpc}$ by the LIGO/Virgo network and to about
 $180\, \mathrm{kpc}$ with the Einstein Telescope.
\end{abstract}

\begin{keywords}
gravitational waves -- supernovae 
\end{keywords}

\section{Introduction}
\label{sec:intro}

The gravitational wave signal emitted by a core-collapse supernova
(CCSN) explosion is expected to be in the detectable frequency range
of ground based gravitational wave detectors such as Advanced LIGO
(aLIGO; \citealp{aLIGO}) and Advanced Virgo (AdVirgo;
\citealp{AdVirgo}). At present, the explosion mechanism of CCSNe is
still not fully understood. The shock wave formed during the core
bounce stalls at a radius of $\sim150\,\mathrm{km}$ and energy is
needed to revive the shock. The current prevailing theory is known as
the neutrino-driven mechanism \citep{2012ARNPS..62..407J,
  2013RvMP...85..245B}, which involves the re-absorption of a fraction
of the emitted neutrinos to heat the post-shock matter, revive the
shock, and power the explosion.

Accurate three-dimensional (3D) CCSN simulations will be essential for understanding the
expected gravitational wave signal and will improve the prospects for
detection and astrophysical interpretation of the
source. Understanding the time-frequency structure will allow tuning
of gravitational wave searches and parameter estimation
\citep{2018MNRAS.474.5272T, 2017PhRvD..96l3013P, 2016PhRvD..94l3012P,
  2018arXiv180207255G, 2016PhRvD..93d2002G}.  The structure of the
gravitational wave signal and the problem of parameter estimation has
already been very thoroughly investigated in the case of the rotational
bounce signal \citep[see][and references
  therein]{2008PhRvD..78f4056D,2014PhRvD..90d4001A,kotake_13,kotake_handbook}.
In this case, the amplitude and frequency can be directly related to
the rotational properties of the core \citep{2008PhRvD..78f4056D,
  2014PhRvD..90d4001A, 2017PhRvD..95f3019R, 2010A&A...514A..51S} and
the frequency of the fundamental quadrupole mode of the proto-neutron
star (PNS) \citep{fuller_15}, which depends on the nuclear equation of
state. However, rapidly-rotating progenitors are expected to occur in
only a small fraction of all CCSNe \citep{2012Natur.481...55B,
  2012A&A...548A..10M}.

For a generic CCSN, one expects the gravitational wave emission
to be dominated by hydrodynamic instabilities in the post-bounce phase
such as turbulent convection in the neutrino heating region and inside the
PNS \citep{herant_94,burrows_95,janka_95}, and the standing accretion
shock instability (SASI; \citealp{0004-637X-584-2-971,
  2006ApJ...642..401B, 2007ApJ...654.1006F}). The gravitational wave
signal from these instabilities has been thoroughly investigated in
axisymmetric (2D) models
\citep{2009ApJ...707.1173M,yakunin_10,2013ApJ...766...43M,yakunin_16,morozova_18},
which have established that the most dominant feature in the
time-frequency structure of the signal is the quadrupolar surface
g-mode of the PNS, which is excited by the hydrodynamical
instabilities inside and outside the PNS. The gravitational wave
emission from the surface g-modes is emitted at high frequencies
of several hundred Hz and rises monotonically in time due to the
contraction of the PNS
\citep{2009ApJ...707.1173M,2013ApJ...766...43M,2013ApJ...779L..18C,
  2017MNRAS.468.2032A, 2016ApJ...829L..14K}; it may thus allow us to
infer properties of the PNS without fully understanding the expected
time series of CCSN gravitational wave signals
\citep{2018MNRAS.474.5272T}.

The gravitational waveforms for the post-bounce signal from 2D
simulations are problematic, however, since significant differences
have been identified between 2D and 3D models in the
post-bounce phase \citep{2017MNRAS.468.2032A}. 
Some gravitational wave signal predictions from 3D simulations of neutrino-driven
explosions have recently become available \citep{2017MNRAS.468.2032A,
  2017arXiv170107325Y, 2018MNRAS.475L..91T, 2017arXiv170107325Y,
  2016ApJ...829L..14K, 2018arXiv181007638A}. The high-frequency
signal from the surface g-mode appears considerably weaker in
3D \citep{2017MNRAS.468.2032A}. In addition, the latest 3D simulations
find that the SASI produces features in the gravitational wave signal at
low frequencies (below 200\,Hz), which is the most sensitive frequency range
of current ground based gravitational wave detectors.

The corpus of gravitational waveforms from 3D simulations of the
post-bounce phase is still small, however, and a number of better
waveforms are still needed for many reasons: Both parameterised 3D
models \citep{2012A&A...537A..63M} and self-consistent 2D models
\citep{2013ApJ...766...43M} have shown that gravitational wave
emission from neutrino-driven convection peaks after shock revival,
yet the signal predictions from self-consistent, state-of-the-art 3D
simulations \citep{2017arXiv170107325Y, 2017MNRAS.468.2032A,
  2018arXiv181007638A} do not reach far into the explosion phase, and
may thus underestimate the peak amplitudes and total power of the
signal.  Moreover, the predicted amplitudes and the characteristic
frequencies differ widely; for example \citet{2017arXiv170107325Y}
find considerably larger amplitudes than \citet{2017MNRAS.468.2032A}
and \citet{oconnor_18}. The time-frequency structure of some models
\citep{2017MNRAS.468.2032A} is not completely clear because of
aliasing problems.  Some waveforms in \citet{2017MNRAS.468.2032A} and
\citet{oconnor_18} are from non-exploding models and may therefore
have unrealistically weak gravitational wave emission. Finally, many
3D simulations use pseudo-Newtonian codes that systematically
overpredict gravitational wave frequencies
\citep{2013ApJ...766...43M}.  We therefore seek to generate new
waveforms from general-relativistic 3D long-time explosion models with
fine time resolution across the mass range of supernova progenitors.

In this study, we perform 3D simulations of two different progenitors
to this end, and analyse the detectability of the predicted waveforms
by current and third-generation gravitational wave detectors. The
first progenitor is an ultra-stripped star simulated from an initial
helium star of mass $3.5\,M_{\odot}$ in \citet{2015MNRAS.451.2123T},
which we refer to as model He3.5.  An ultra-stripped star is a binary star that has
become an almost naked metal core due to mass transfer via 
Roche-lobe overflow to the binary companion
\citep{2015MNRAS.451.2123T, 2013ApJ...778L..23T}. Although supernovae from
stripped-envelope progenitors in binary systems are of interest in
their own right because most massive stars are located in binaries
\citep{2012MNRAS.424.1925C} and a large fraction of them interact with
their companion \citep{2013ASPC..470..141S}, this model primarily
serves as a representative example for a supernova with modest
explosion energy \citep{2018arXiv181105483M}. The second progenitor is
an $18\,M_{\odot}$ star.  This more massive progenitor is
representative for supernovae with higher, more typical explosion
energies around $10^{51}\,\mathrm{erg}$ and stronger
gravitational wave emission. Since more energetic supernovae are
expected to have stronger gravitational wave emission
\citep{2013ApJ...766...43M}, these two progenitors allow us to better
assess the range of gravitational wave amplitudes from successful
explosion models. Aside from producing gravitational waveforms that
reach further into the explosion phase, another thrust of our paper is
to shed more light on the dynamics of the PNS convection, which has
been found to be crucial for the excitation of the PNS surface g-mode
in 3D \citep{2017MNRAS.468.2032A}, and whose relation to the
lepton-number emission self-sustained asymmetry (LESA;
\citealp{tamborra_14}) is still not well understood.

The outline of our paper is as follows: Section~\ref{sec:sim} gives a
brief outline of our numerical methods and set up. In
Section~\ref{sec:models}, we describe the models that are simulated in
this study. In Section~\ref{sec:dynamics}, we discuss the explosion
dynamics of our models and compare to previous simulations. In
Section~\ref{sec:pns_conv}, we describe and analyse the dynamics of
the PNS convection zone and discuss the features of the LESA
instability observed in our models. The features of the gravitational
wave emission are described in Section~\ref{sec:gw}. We explore the
detectability of our gravitational wave signals in
Section~\ref{sec:detection}, and a discussion and conclusions are
given in Section~\ref{sec:conclusion}.

\section{Simulation Methodology and Setup}
\label{sec:sim}

Our simulations are carried out using the neutrino hydrodynamics code
\textsc{CoCoNuT-FMT}. We use a general relativistic finite-volume
based solver for the equations of hydrodynamics
\citep{2010ApJS..189..104M, 2002A&A...393..523D} formulated in
spherical polar coordinates. Different from previous 3D simulations
with \textsc{CoCoNuT-FMT}, we simulate in 3D down to the innermost
$\sim10$\,km to include the PNS convection zone and impose spherical
symmetry inside this radius. The neutrino transport is handled using
the fast multigroup transport method of \citet{mueller_15a} with
updates to the neutrino rates to include nucleon potentials
\citep{martinez_12}, nucleon correlations in the virial approximation
\citep{horowitz_17}, weak magnetism \citep{horowitz_02}, and a nucleon
strangeness of $g_A=-0.05$, which is roughly compatible with current
experimental constraints and theoretical expectations \citep{hobbs_16}.
To keep the computational costs manageable despite resolving the
PNS convection zone in 3D, we adopt a coarser grid in energy
space with only 9 energy groups (instead of the usual 21 energy
groups in our recent models).

We implemented on-the-fly extraction of gravitational waves with fine
temporal sampling, by means of the modified time-integrated quadrupole
formula for strong fields as described in \citet{2013ApJ...766...43M}.

At high density, we use the equation of state from \citet{Lattimer:1991nc} with a bulk incompressibility of K=220\,MeV, and the standard treatment of the low-density regime in \textsc{CoCoNuT-FMT} with
an equation of state for photons, electrons, positrons, and an ideal gas of nuclei together with
a flashing treatment for nuclear reactions \citep{rampp_02}.

\section{Supernova Models}
\label{sec:models}

We simulate the explosion of two different models, namely model
He3.5, an ultra-stripped star \citep{2015MNRAS.451.2123T} with a modest
explosion energy of $\mathord{\sim}0.3$\,Bethe \citep{2018arXiv181105483M}, and model
s18, an $18\,M_{\odot}$ single-star progenitor with a higher, more typical explosion
energy of $\mathord{\sim}0.8$\,Bethe \citep{2017MNRAS.472..491M}. We refer to the
original simulations of \citet{2017MNRAS.472..491M} and \citet{2018arXiv181105483M},
as s18-old and He3.5-old to differentiate them from the new models
calculated with the setup described in Section~\ref{sec:sim}.

\subsection{Model He3.5}

Model He3.5 is an ultra-stripped star that was evolved from a helium star
with an initial mass of $3.5\,M_{\odot}$ 
\citep{2015MNRAS.451.2123T}. The model was evolved with the binary
evolution code BEC \citep{2001A&A...369..939W} to simulate the later
phases of mass loss in the ultra-stripped supernova channel. The pre-collapse mass
of the star is $2.6\,M_{\odot}$. Our simulations cover the region
inside $10^{5}\,\mathrm{km}$, as the envelope outside this region does not
change on the timescale of seconds. During the collapse and the
early post-bounce phase, we simulate the interior of the PNS
at densities $\mathord{>}10^{11}\, \mathrm{g}\, \mathrm{cm}^3$
in spherical symmetry using a mixing-length treatment
as in \citet{mueller_15b}; up to that point, no gravitational wave
amplitudes are computed. The region outside is always simulated
in 3D. We extend the 3D region to include the
PNS convection zone and reduce the spherically symmetric inner domain
to a radius of $10\, \mathrm{km}$ when we restart the simulation 0.05\,s after core bounce.
The simulation is stopped at 0.7\,s after core bounce.
Aside from the lower energy resolution, the only other difference
between model He3.5 and He3.5-old is the inclusion of weak magnetism
in He3.5.

\subsection{Model s18}

Model s18 is a solar metallicity progenitor star with a zero-age main
sequence (ZAMS) mass of $18\,M_{\odot}$, and a helium core mass of
$5.3\,M_{\odot}$. The last minutes of
oxygen shell burning were simulated in 3D \citep{2016ApJ...833..124M} to properly
initialise the aspherical seed perturbations that are critical for shock revival
for this model \citep{2017MNRAS.472..491M}. Although model s18 has
been simulated previously in \citet{2017MNRAS.472..491M}, the
gravitational wave signal was not calculated as the PNS convection
zone, which is the main emission region for gravitational waves, was
simulated in spherical symmetry.
We restart the simulation with a full 3D treatment of
PNS convection 0.08\,s after the core bounce and stop the simulation at 0.89\,s after core bounce. 
Model s18 also differs from s18-old in that we include strangeness corrections, nucleon correlations,
and weak magnetism.

\begin{figure*}
\centering
\includegraphics[width=\columnwidth]{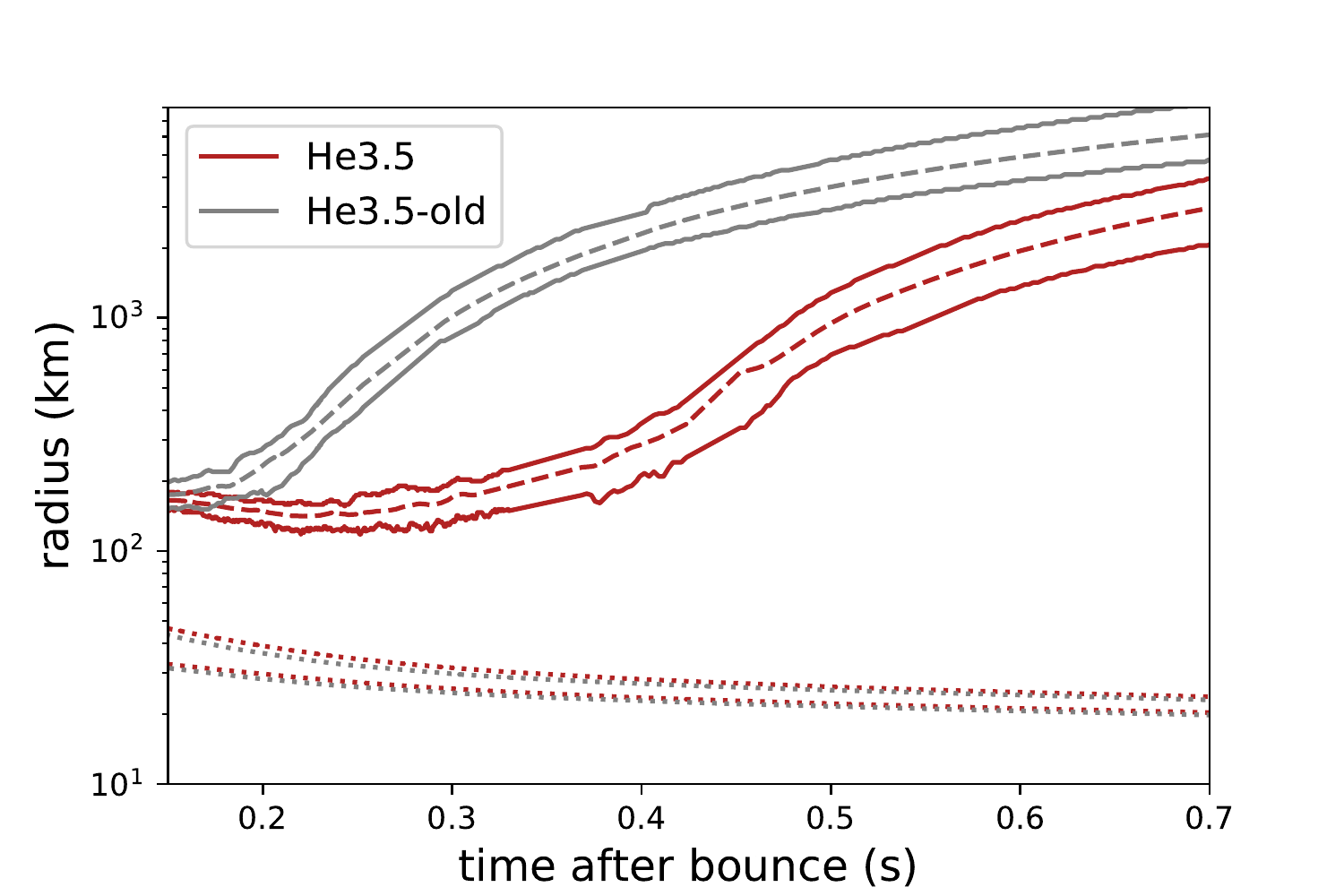}
\includegraphics[width=\columnwidth]{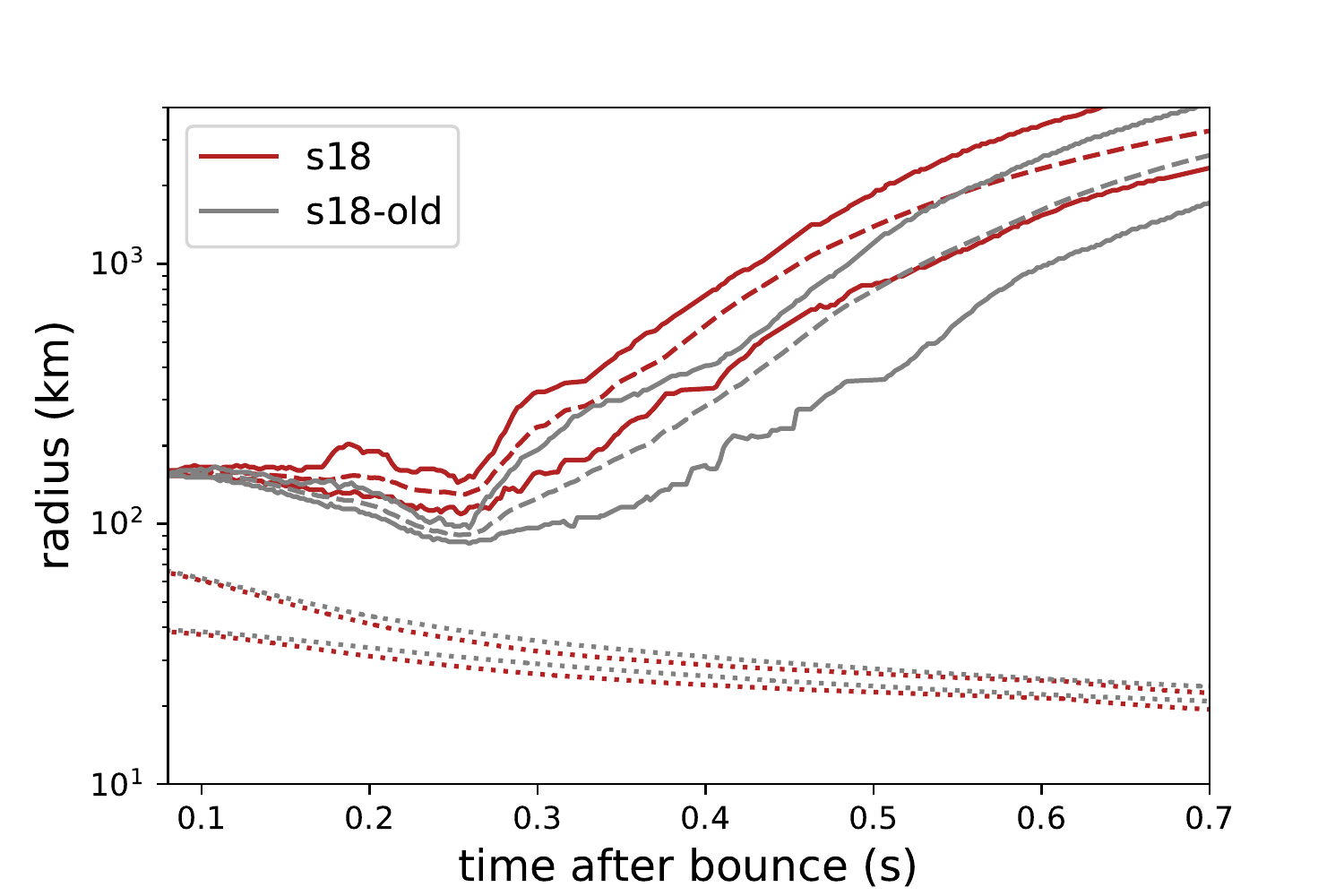}
\caption{The minimum, maximum, and average shock radius (solid lines) and the radii corresponding to densities of $10^{11}\mathrm{g}\,\mathrm{cm}^{-3}$ and $10^{12}\mathrm{g}\,\mathrm{cm}^{-3}$ (dotted lines, tracking the contraction of the PNS surface) for model s18 (right) and model He3.5 (left). The gray lines show the results from previous simulations, and the red lines show the results from our models. Model He3.5 explodes much later than the previous simulation and has a smaller shock radius. Model s18 explodes slightly earlier than the previous simulation and has a larger shock radius.}
\label{fig:radius}
\end{figure*}

\begin{figure*}
\centering
\includegraphics[width=\columnwidth]{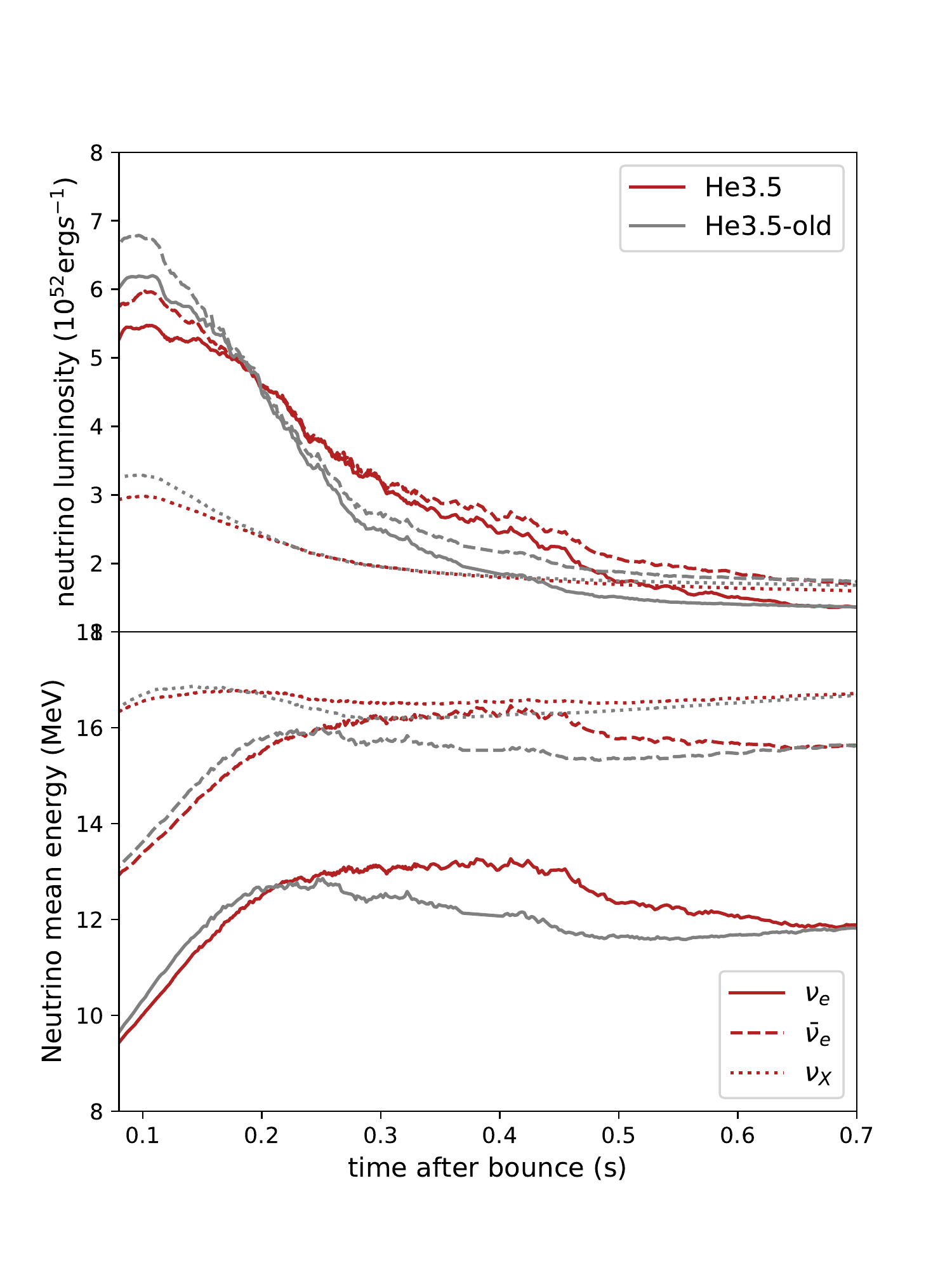}
\includegraphics[width=\columnwidth]{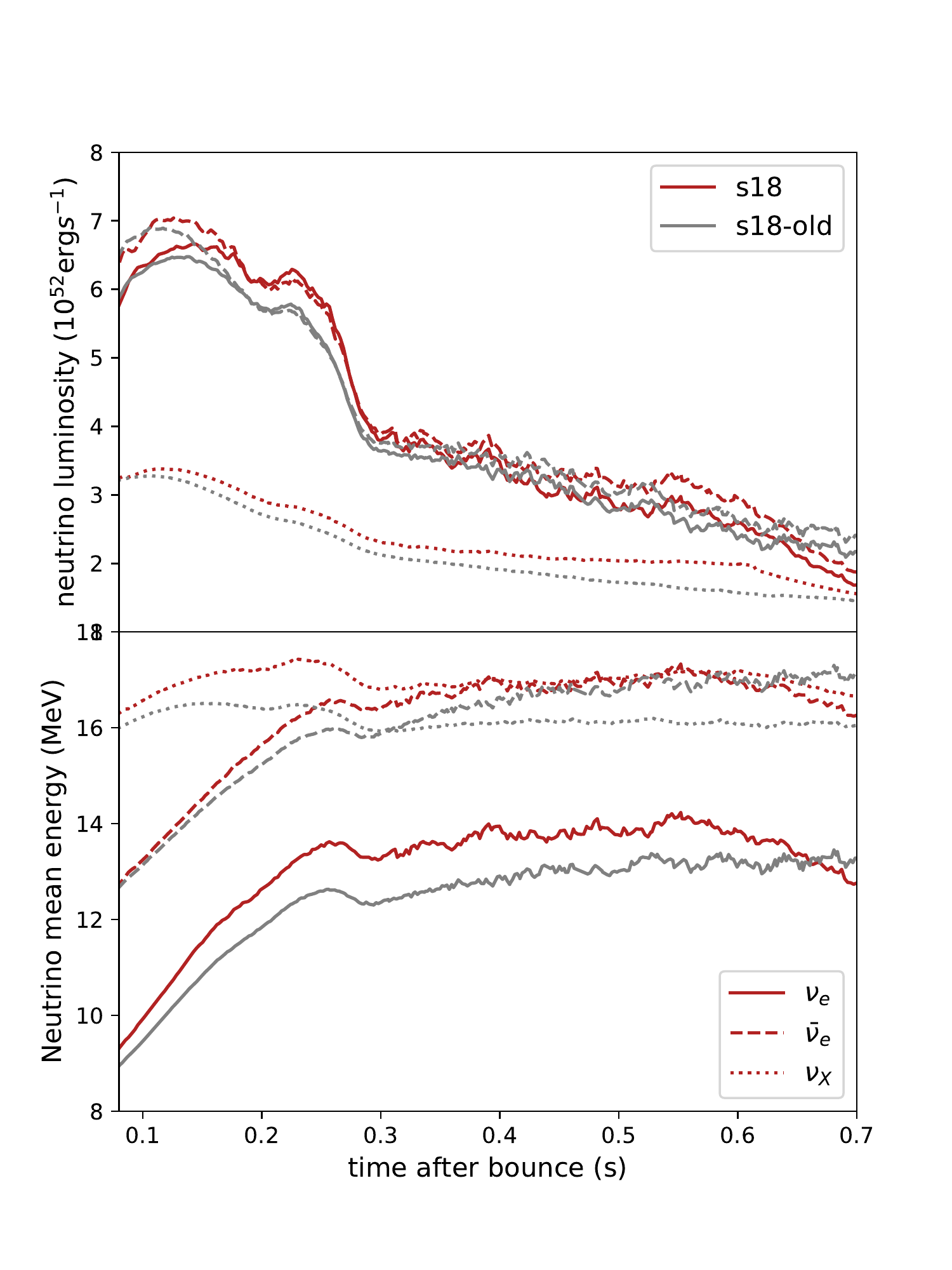}
\caption{Neutrino luminosities (top) and mean energies (bottom) of $\nu_e$ (solid), $\bar{\nu}_e$ (dashed), and heavy flavour neutrinos $\nu_X$ (dotted) for model s18 (right) and model He3.5 (left) compared
to the published models s18-old and He3.5-old. The gray lines show the results from previous simulations, and the red lines show the results from our models. Model He3.5 has lower neutrino luminosity than He3.5-old within the first 0.2\,s, and higher neutrino luminosity at later times than the previous simulation. Model s18 has higher neutrino energies than model s18-old. 
\label{fig:neutrino}}
\end{figure*}

\section{Explosion Dynamics and Comparison with Previous Simulations}
\label{sec:dynamics}

Even though the two progenitors have already been simulated in 3D with the \textsc{CoCoNuT-FMT} code, it is useful to briefly
outline the dynamical evolution of both models as background information for understanding the gravitational wave signals.
Moreover, since we changed several elements of the simulation setup,
we briefly compare our models to the corresponding simulations of \citet{2018arXiv181105483M} and \citet{2017MNRAS.472..491M}. 
Although we cannot disentangle the individual effects of the full 3D treatment of PNS convection, the improved neutrino opacities,
and the reduced resolution in energy space, it is important to verify that the new simulations remain qualitatively and
quantitatively similar to the old models. It is especially important that the physical parameters affecting the gravitational wave
emission are not unduly affected by the reduced energy resolution.

The shock radius and the radii corresponding to densities of $10^{11}\mathrm{g}\,\mathrm{cm}^{-3}$ and $10^{12}\mathrm{g\,cm}^{-3}$ are shown 
in Figure \ref{fig:radius} for both models. 
Models He3.5 and He3.5-old both explode due to the relatively steep decline of the mass accretion rate that is
characteristic of stars with low-mass helium cores. However, the shock is revived considerably
\emph{later} in model He3.5 than in He3.5-old, and the shock radius remains significantly smaller from 0.2\,s after core bounce onward. PNS contraction is slightly slower in He3.5 than in He3.5-old.

Model s18 undergoes explosion $\mathord{\sim}250\,\mathrm{ms}$ after bounce aided by the infall of strong seed perturbations from
the O shell. The simulation produced in this paper shows a slightly larger shock radius than s18-old, and the shock is revived $\mathord{\sim}5\,\mathrm{ms}$ earlier. The PNS contracts slightly faster in s18 than in s18-old.

The  differences in the evolution of model s18 vis \`a vis s18-old and He3.5 vis \`a vis He3.5-old are explained by
differences in the neutrino emission that result from the changes in energy resolution and the opacities.
The neutrino luminosities and mean energies  for all models are shown in Figure \ref{fig:neutrino}. For the
$18\,M_\odot$ progenitor, the changes in energy resolution and neutrino interaction rates result in 
increased luminosities and mean energies of all neutrino flavours.
This naturally explains the slightly earlier onset of the explosion and the larger stagnation of the shock radius prior to shock revival.
Since the infall of the Si/O shell interface and the presence of pre-shock perturbations in the O shell play a major role in triggering
shock revival, the higher neutrino luminosities and mean energies cannot shift the onset of the explosion appreciably, however, and hence the shock trajectories of models s18 and s18-old remain similar during the explosion phase. The somewhat faster
propagation of the shock in s18 helps to quench accretion onto the PNS, and as the result of the slightly faster decline
of the accretion luminosity, the neutrino emission in model s18 again becomes more similar to s18-old later in the
explosion phase. 

The increased neutrino luminosities and mean energies in model s18 likely result from a combination of effects.
Both the strangeness correction for neutrino-nucleon scattering
\citep{melson_15} and nucleon correlations \citep{horowitz_17,bollig_17} tend to increase neutrino luminosities
and mean energies and lead to faster PNS contraction.\footnote{The primary effect of the reduced scattering opacities is stronger cooling in heavy flavour neutrinos, which then results in faster PNS contraction. Increased electron flavour luminosities come
about as a secondary effect due to increased temperatures at the respective neutrinospheres \citep{melson_15}.} 
This effect may be compounded by the low energy resolution. Only a few studies have addressed the role of resolution in energy
space, but it is clear that nine energy groups are not sufficient to fully
resolve the neutrino spectrum and easily introduce uncertainties
of a few percent in the luminosities \citep{marek_phd}.

The results for models He3.5 and He3.5-old confirm that energy resolution is a major
factor. Figure~\ref{fig:neutrino} shows a significant reduction of
the neutrino luminosities of all flavours 
on the 10\%-level in He3.5-old early on. Although model He3.5 also
differs from He3.5-old in that it includes weak magnetism
corrections, these corrections do not produce an effect of such magnitude
\citep{buras_06a,fischer_18}, and do not affect the heavy flavour neutrino in our simulations since we treat $\nu_{\mu/\tau}$ and their antiparticles $\bar{\nu}_{\mu/\tau}$ as a single species in our code.

Due to the relatively large differences in the neutrino emission, shock
revival is delayed by about $0.2 \, \mathrm{s}$ in model He3.5
compared to He3.5-old. As a result of ongoing accretion, higher
neutrino luminosities and mean energies are maintained in model
He3.5 from a post-bounce time of $0.2 \, \mathrm{s}$ until
about $0.65\, \mathrm{s}$. Moreover, the PNS radius is slightly larger in model He3.5.

From this cursory analysis, it is clear that the energy resolution and the details
of the microphysics have an important bearing on the detailed neutrino emission.
This is, however, not our primary concern here. What is important for our purpose is rather whether the models predict the physical parameters relevant for the gravitational wave emission with reasonable robustness. Especially
for model s18, this appears to be the case: Despite differences in energy
resolution and in the microphysics, models s18 and s18-old exhibit very
similar explosion dynamics. The factors that determine the frequency
$f_\mathrm{g}$ of the surface g-mode that dominates
the gravitational wave spectrum \citep{2009ApJ...707.1173M,2013ApJ...766...43M} are also similar.
In terms of the baryonic neutron star mass $M_\mathrm{by}$,
the neutron star radius $R$,  the electron antineutrino mean
energy $\langle E_{\bar{\nu}_\mathrm{e}}\rangle$,
and the neutron mass $m_n$ one
one finds \citep{2013ApJ...766...43M},
\begin{equation}
f_\mathrm{g}\approx \frac{1}{2\pi}\frac{GM_\mathrm{by}}{R^2}
\sqrt{1.1\frac{m_\mathrm{n}}{\langle E_{\bar{\nu}_\mathrm{e}}\rangle}}
\left(1-\frac{GM_\mathrm{by}}{Rc^2}\right).
\end{equation}
As discussed before, $\langle E_{\bar{\nu}_\mathrm{e}}\rangle$
and $R$ do not differ greatly between models s18 and s18-old, and
the PNS masses are also similar; the
final baryonic PNS mass $M_\mathrm{by}$ is lower by $0.06\,M_\odot$ in model s18 due to somewhat faster quenching of the accretion.
Hence we expect the frequency of the surface g-mode to be reliable
within a few percent despite the reduced energy resolution in model s18.

The case of the $3.5\,M_\odot$  stripped-envelope progenitor is slightly different
because the explosion dynamics of model He3.5 differs appreciably from
He3.5-old. When interpreting the gravitational wave amplitudes, it
must be borne in mind that there is considerably uncertainty about the
time of shock revival in this model. There is less
of a concern about the gravitational wave spectra, however. Although
the neutrino mean energies differ by up to $\mathord{\sim}10 \%$
between He3.5 and He3.5-old (as a result of the delayed onset
of the explosion), the PNS contraction is similar, as is the PNS
mass, which is only $0.03\,M_\odot$ higher in He3.5 at the end of
the simulation. Since the electron antineutrino mean energy
only enters in $f_\mathrm{g}$ as $\langle E_{\bar{\nu}_\mathrm{e}}\rangle^{-1/2}$, the surface g-mode frequency in model He3.5 is
still reliable within a few percent.

\begin{figure}
  \centering
\includegraphics[width=\columnwidth]{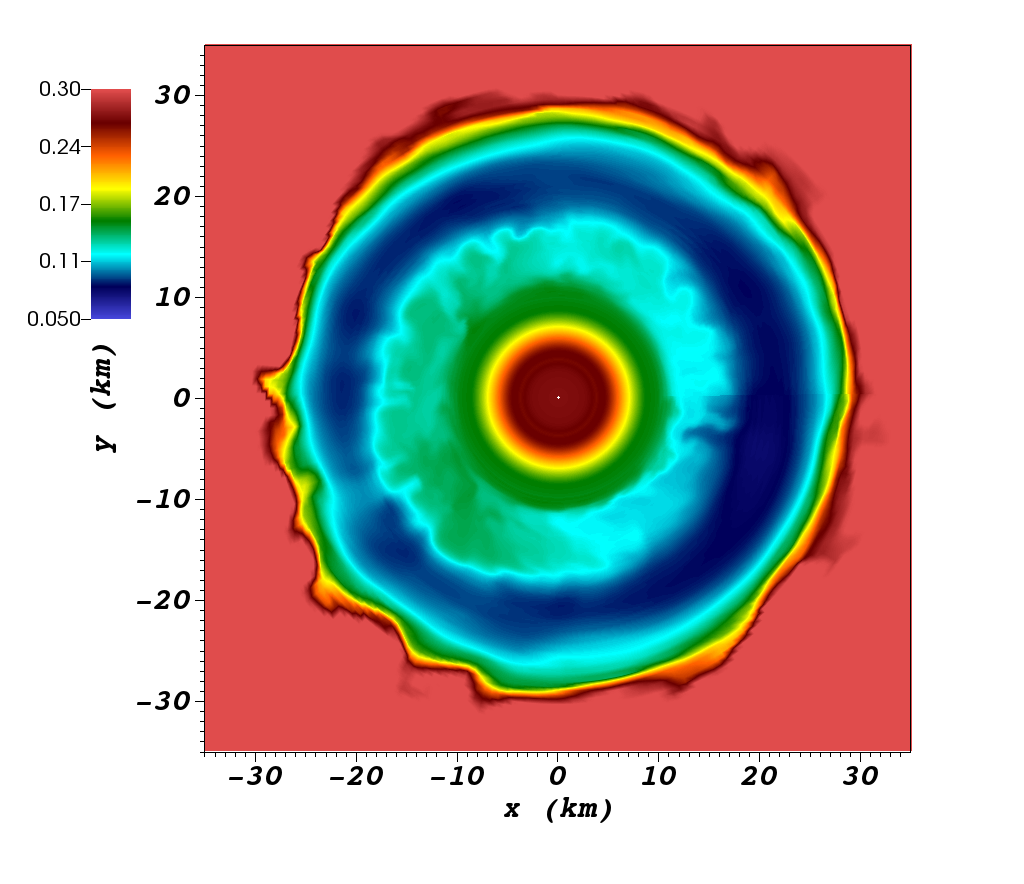}\\
\includegraphics[width=\columnwidth]{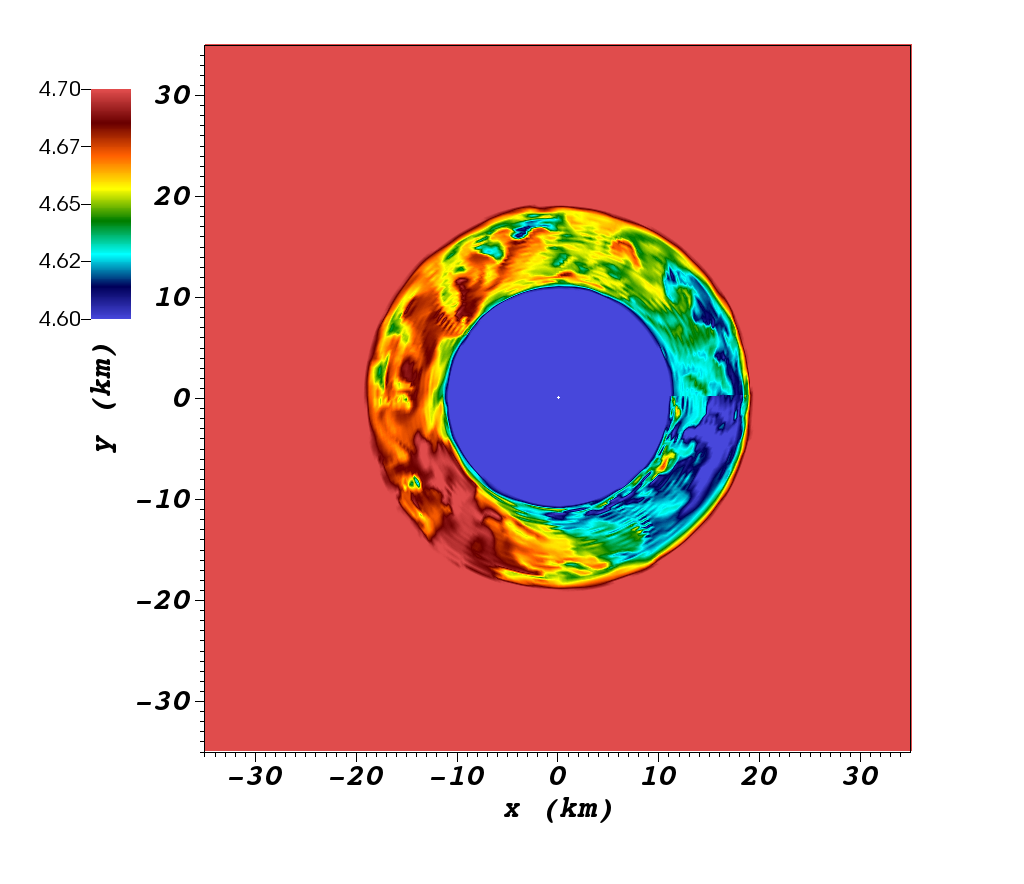}\\
\includegraphics[width=\columnwidth]{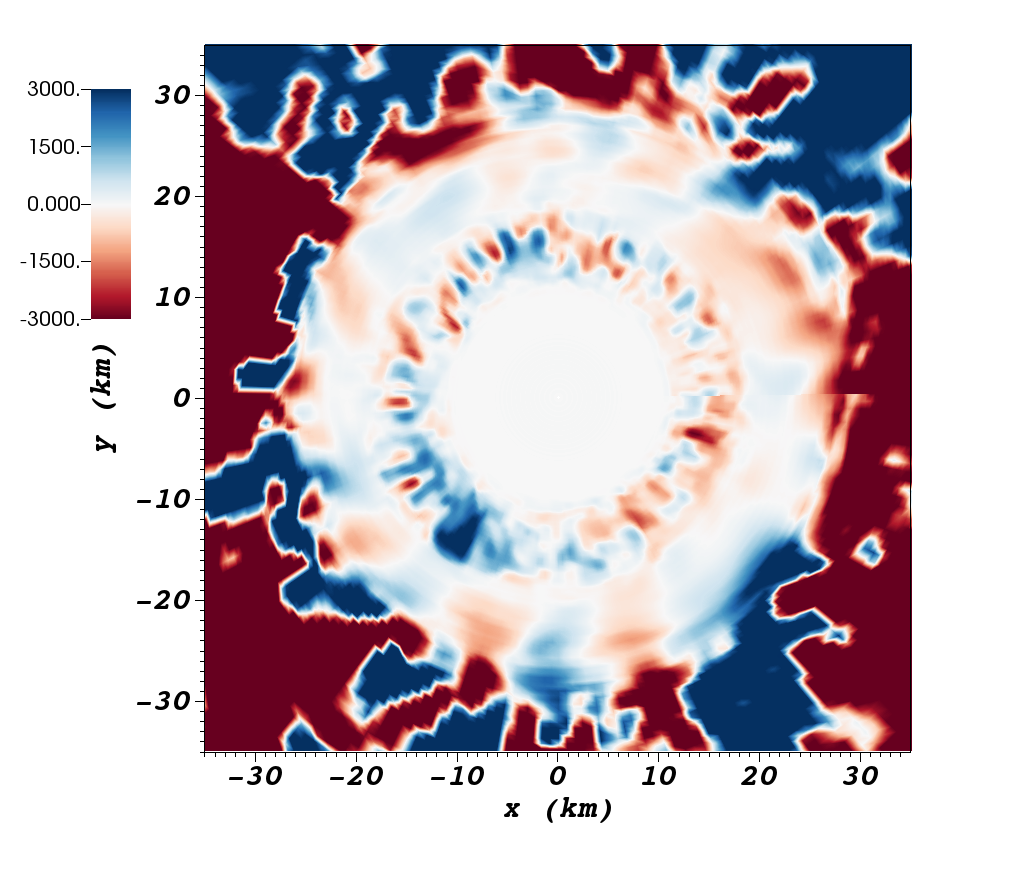}\\
\caption{Two-dimensional slices through the PNS convection zone
  in model s18 showing the electron fraction (top),
  the entropy in units of $k_\mathrm{b}/\mathrm{nucleon}$
  (middle), and the radial velocity in units of $\mathrm{km}\, \mathrm{s}^{-1}$
  at a post-bounce time of $453\, \mathrm{ms}$.
  \label{fig:pns_convection}}
\end{figure}

\begin{figure}
  \centering
  \includegraphics[width=\columnwidth]{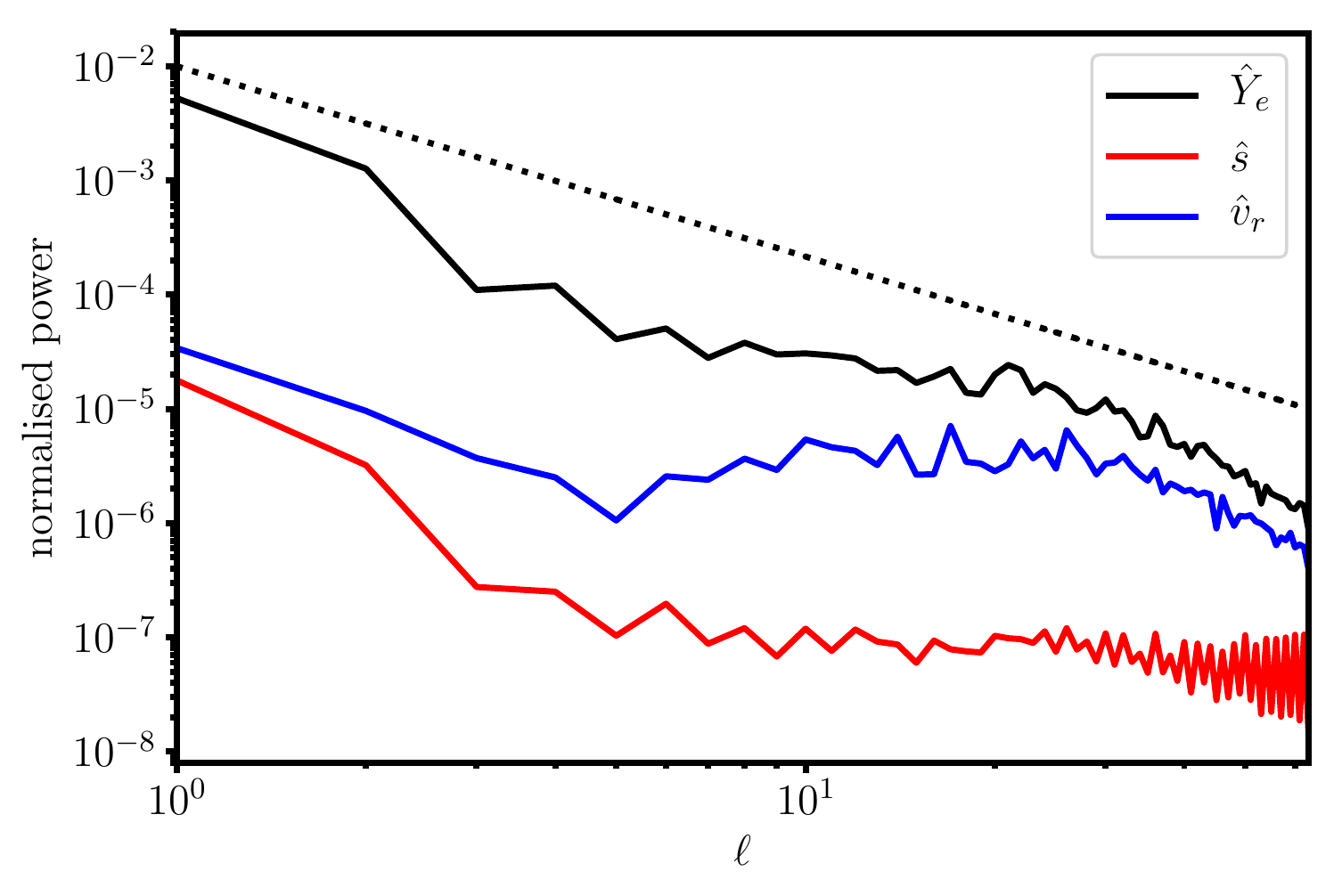}\\
  \caption{Spectra of the normalised perturbations in electron
    fraction ($\hat{Y}_{e,\ell}/\hat{Y}_{e,0}$, black), entropy
    ($\hat{s}_{e,\ell}/\hat{s}_{e,0}$, black) and the radial Mach
    number ($\hat{v}_{r,\ell}/\hat{c}_{s,0}$, blue) in the PNS
    convection zone at a radius of $13.7\, \mathrm{km}$ and at a
    post-bounce time of $454 \, \mathrm{ms}$ for model s18.  The slope of a
    Kolmogorov spectrum with a power law of $-5/3$ is indicated as a
    dotted line.
  \label{fig:pns_spec}}
\end{figure}

\begin{figure}
  \centering
  \includegraphics[width=\columnwidth]{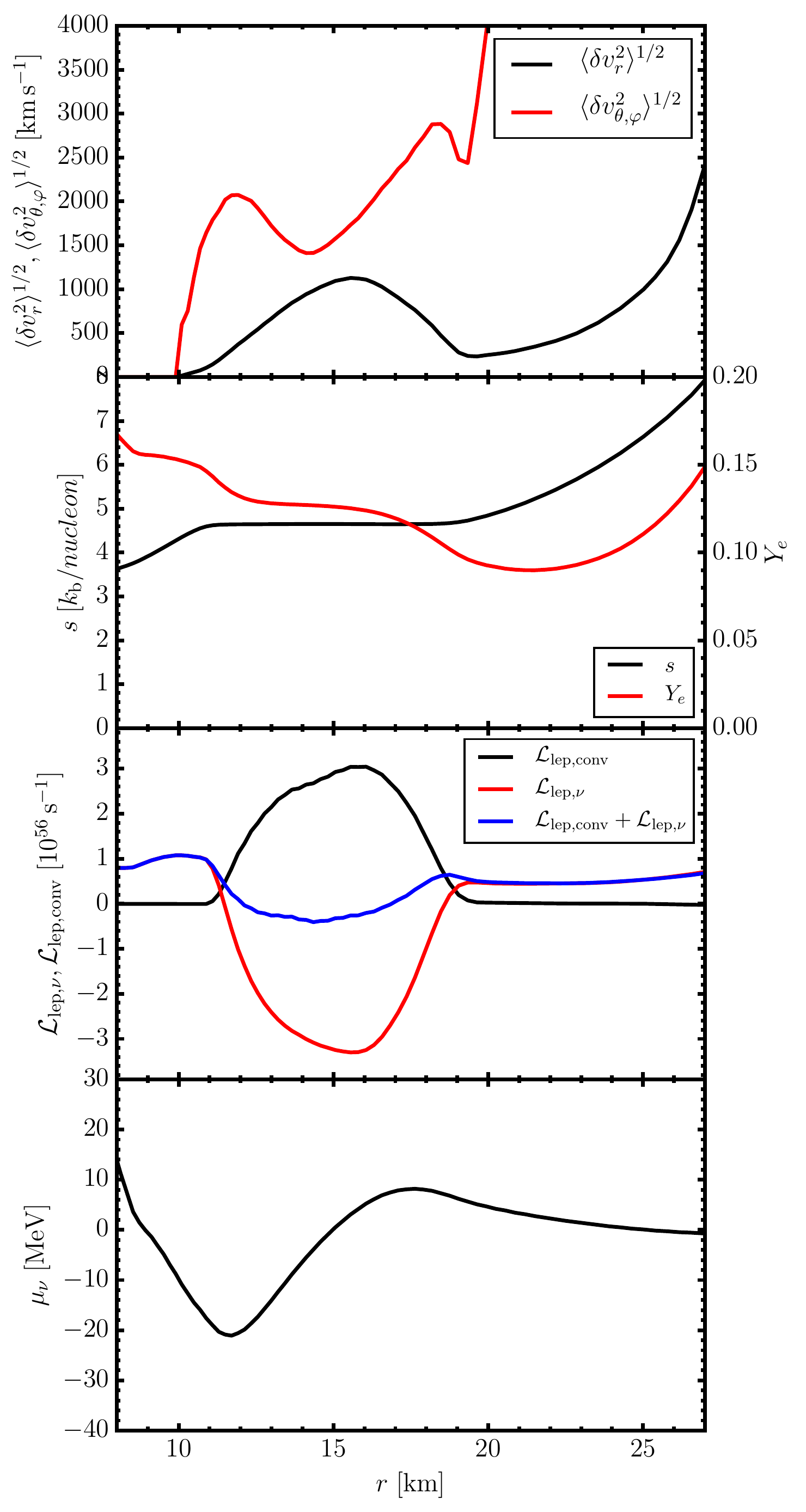}\\
  \caption{
    RMS averages of the radial and non-radial
    velocity fluctuations (panel 1), angle-averaged
    profiles of the entropy $s$ and the electron fraction 
    $Y_e$ (panel 2), convective and diffusive
    lepton number luminosities
    $\mathcal{L}_\mathrm{\mathrm{lep},\mathrm{conv}}$ 
    and $\mathcal{L}_\mathrm{\mathrm{lep},\nu}$ and their sum (panel 3),
    and neutrino chemical potential $\mu_\nu$ (panel 4)
    in the PNS convection zone for model s18 at
    a post-bounce time of $454\, \mathrm{ms}$.
  \label{fig:profile}}
\end{figure}

\begin{figure}
  \centering
\includegraphics[width=\columnwidth]{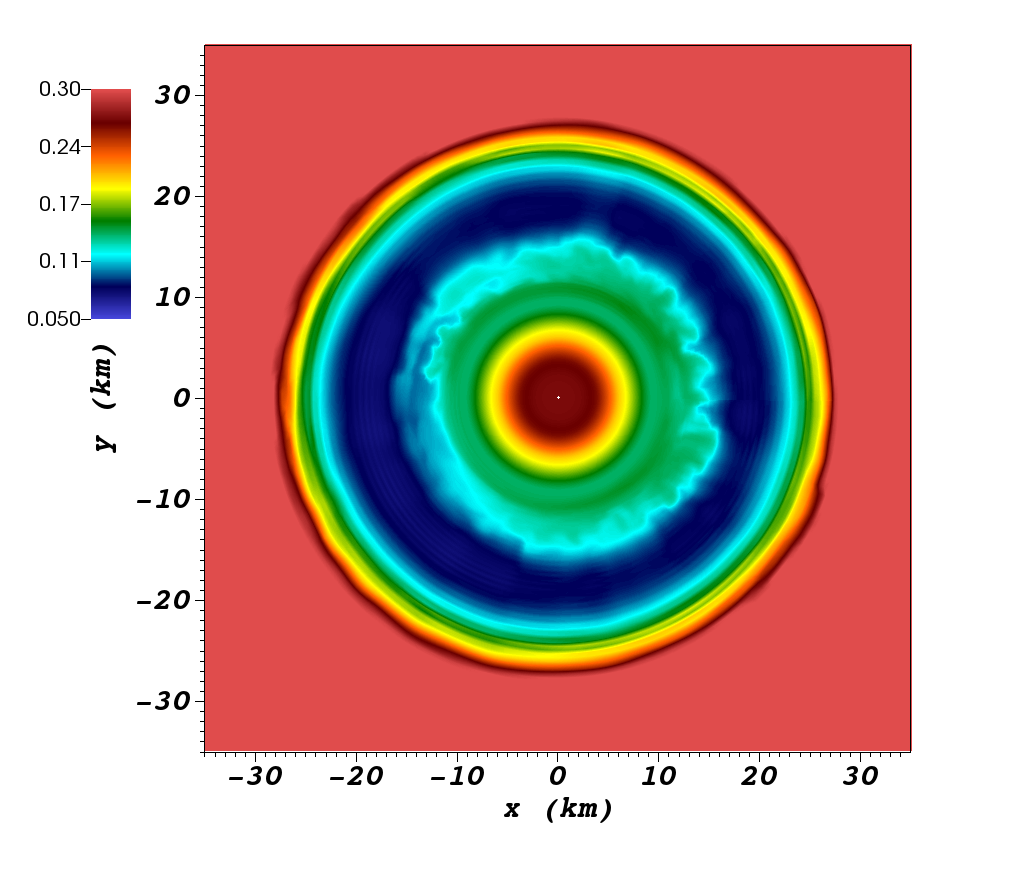}\\
\includegraphics[width=\columnwidth]{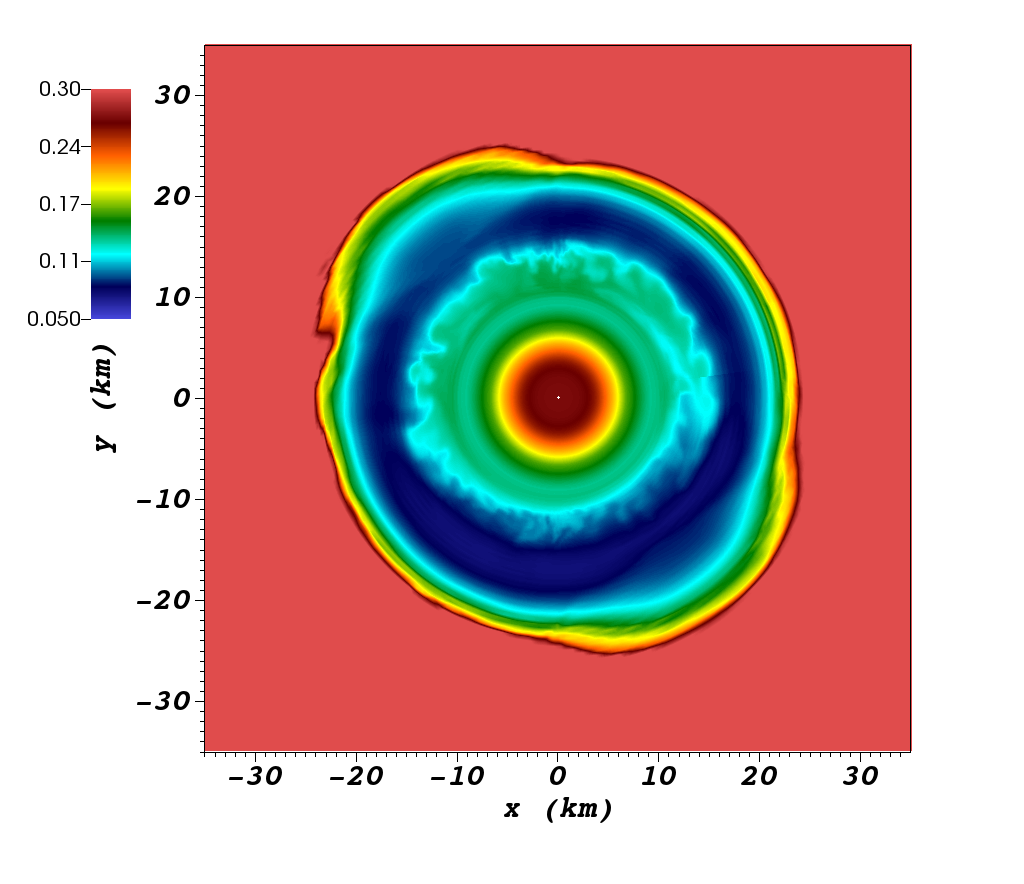}\\
\caption{Two-dimensional slices through the PNS convection zone
  showing the electron fraction 
  in models s3.5 (top) and
  s18 (bottom)
  towards the end of the simulations.
  \label{fig:lesa_final}}
\end{figure}

\section{Features of Proto-Neutron Star Convection}
\label{sec:pns_conv}

In their study of gravitational wave emission in 3D supernova models,
\cite{2017MNRAS.468.2032A} concluded that PNS convection is the
primary driver of the PNS surface g-modes that dominate the
high-frequency gravitational wave spectrum. Although multi-dimensional
simulations of PNS convection initially suggested a relatively
straightforward picture of PNS convection as simple Ledoux convection
\citep{keil_96,buras_06b,dessart_06}, the fluid motions in the PNS
convection zone have proved more enigmatic in recent years.  While the
extent of the unstable region is well described by the Ledoux
criterion and quite consistent between different codes, significant
variations in the convective velocities have recently been reported in
a cross-comparisons of the \textsc{Vertex} and \textsc{Alcar} codes
\citep{just_18}. Moreover, many of the recent 3D models develop a
peculiar hemispheric lepton number asymmetry in the PNS convection
zone as first discovered and termed LESA (``Lepton-number Emission
Self-sustained Asymmetry'') instability by \citet{tamborra_14}; this
has recently been confirmed in other 3D simulations using different
codes \citep{oconnor_18,glas_18}. Independently of computational
studies of PNS convection, the role of double-diffusive instabilities \citep{bruenn_96}
in the PNS convection zone has long been discussed, but no phenomenon
in multi-D simulations has yet been connected to such instabilities.
What precisely determines the flow pattern and the convective
velocities in the PNS interior is thus by no means completely understood.

Before we proceed to analysing the gravitational wave emission itself,
it is therefore worthwhile to study the properties of PNS convection
in our 3D models and compare the findings to other recent
simulations.
Since all of the qualitative features of PNS
convection turn out to be in agreement with the studies
of \citet{tamborra_14,oconnor_18,glas_18}, we
mostly confine our analysis
to model s18 and to one representative epoch.

Figure~\ref{fig:pns_convection} shows the electron fraction $Y_e$, the
specific entropy $s$, and the radial velocity $v_r$ on two-dimensional
slices through model s18 at a post-bounce time of $454\,
\mathrm{ms}$. LESA is clearly present, with a pronounced dipolar
asymmetry in the electron fraction $Y_e$ (top panel). While previous
studies of LESA have not addressed asymmetries in the entropy (which
are somewhat hard to diagnose because of the minute dynamic range of
entropy fluctuations), the entropy distribution in the PNS convection
zone exhibits the same dipolar asymmetry in our model. This suggests
that LESA is actually not an instability that imprints a global
asymmetry specifically on the lepton number distribution. The lepton
number asymmetry is merely more prominent because the contrast in
electron fraction/lepton number across the PNS convection zone is
considerably larger than the entropy contrast. This could suggest that
LESA is in effect not a new instability, but merely a manifestation of
``normal'' convection with the preferred length scales set by the
geometry of the unstable region and perhaps finite (physical and
numerical) viscosity and diffusivity \citep{glas_18}.  The presence of
weak dipole asymmetry in the radial velocity field (again,
similar to \citealt{glas_18}), superimposed over
eddies of smaller scale  (bottom panel), would be compatible
with this parsimonious interpretation.

A more detailed analysis reveals more complicated dynamics
in the PNS convection zone, however. To analyse the spectrum
of the convective motions, we consider the power
in spherical harmonics of degree $\ell$
for the perturbations in electron fraction, entropy,
and radial velocity. For quantity $X$, we compute
the power $\hat{X}_\ell$ in multipole $\ell$ as 
\begin{equation}
  \hat{X}_\ell=
  \sum_{m=-\ell}^\ell
  \left|\int Y_{\ell m}^*(\theta,\varphi) X(\theta,\varphi) \,\ud \Omega\right|^2.
\end{equation}
It is then useful to normalise the power by the power $\hat{X}_0$ in
the monopole, or, in case of the radial velocity, by the monopole
component $\hat{c}_{\mathrm{s},0}$ of the sound speed.
When computing the spectra, we average the power over six radial zones
around a radius of $13.7\, \mathrm{km}$.

The normalised spectra $\hat{Y}_{e,\ell}/\hat{Y}_{e,0}$,
$\hat{s}_{e,\ell}/\hat{s}_{e,0}$ and $\hat{v}_{r,\ell}/\hat{c}_{s,0}$
are shown in Figure~\ref{fig:pns_spec}. The spectra differ
considerably from standard Kolmogorov scaling laws for isotropic
turbulence and also from Bolgiano-Obukhov scaling for stratified ideal
and viscous flow \citep{obukhov_59,kumar_14}.  The spectra of the electron
fraction and entropy do not nicely conform to a power law at low
$\ell$ and are suggestive of a deficit of power at angular wave
numbers $\ell\sim 3\texttt{-}10$.  For the radial velocity, it is even
clearer that there are two peaks at $\ell=1$ and $\ell \sim 10$,
suggesting that the flow is shaped by two distinct scales. It is also
noteworthy that the shape of the velocity spectrum differs from the
spectra of $s$ and $Y_e$. This is understandable for the wavenumbers
that are subject to strong numerical dissipation or (radial) odd-even
noise that manifests itself in the high-wavenumber tail of the entropy
spectrum, but it is not expected for low and intermediate wavenumbers.

More evidence for complex stratification within the
PNS convection zone comes from radial profiles of various
thermodynamic quantities and the turbulent velocity fluctuations, and
from the turbulent lepton number flux (Figure~\ref{fig:profile}).
The angular root-means square averages $\langle \delta v_r^2
\rangle^{1/2}$ and $\langle \delta v_{\theta,\varphi}^2\rangle ^{1/2}$
of the radial and non-radial velocity fluctuations $\delta v_r$ and
$\delta v_{\theta,\varphi}$ around their spherical average (top panel)
suggest that the well-mixed region extends from about
$\mathord{\sim}10\, \mathrm{km}$ to $\mathord{\sim}19\, \mathrm{km}$.
While the entropy profile is almost perfectly flat within this region
(second panel), the profile of $Y_e$ exhibits a much narrower plateau
in the middle of the convective zone, and broad ``steps'' at the outer
and inner convective boundary. Especially at the outer boundary, this
step clearly starts already well within the convection zone. This
$Y_e$-profile is maintained as the result of convective and diffusive
transport counteracting each other. The third panel in
Figure~\ref{fig:profile} illustrates that through most of the PNS
convection zone, the net lepton number luminosity
$\mathcal{L}_{\mathrm{lep},\nu}$ carried by diffusing neutrinos is
actually negative and almost cancels the convective lepton number
luminosity $\mathcal{L}_\mathrm{\mathrm{lep},\mathrm{conv}}$, which we
compute as\footnote{Note that we do not include the net electron
  neutrino fraction when computing the convective lepton
  number flux, since the FMT scheme does not include the advection
  of neutrinos with the fluid.}
\begin{equation}
  \mathcal{L}_\mathrm{\mathrm{lep},\mathrm{conv}}
  = \int \rho W \delta v_r\, \delta Y_e \alpha \phi^4 r^2 \, \ud \Omega.
\end{equation}
Here $\rho$, and $W$ are the baroynic mass
density and the Lorentz factor, $\delta v_r$ and $\delta Y_e$ are the
fluctuations of the radial velocity and electron fraction computed
from a spherical Favre decomposition, and $\alpha$ and $\phi$ are the
lapse function and conformal factor in the xCFC metric.

The negative diffusive lepton number flux is the consequence of a
positive gradient in the neutrino chemical potential $\mu_\nu$ that
develops in a wide layer within the convective region (bottom panel of
Figure~\ref{fig:profile}). 
The positive gradient in $\mu_\nu$ is
also noteworthy because it is a direct reflection of
the well-established finding \citep{lattimer_81,keil_96,bruenn_96,bruenn_04,buras_06b}
that negative lepton number gradients tend to stabilise
the stratification against convection in some
parts of the convective region. \citet{lattimer_81} showed that
the more familiar form of the Ledoux criterion for convective
instability in the presence of entropy and lepton number (electron
fraction)\footnote{Again, we use the electron fraction instead of the
  lepton number for our further analysis, since the FMT transport
  module does not include the advection of neutrinos with the fluid.},
\begin{equation}
  \left( \frac{\pd \rho}{\pd s} \right)_{Y_e,P} \frac{\ud s}{\ud r}
  +\left( \frac{\pd \rho}{\pd Y_e} \right)_{s,P} \frac{\ud Y_e}{\ud
    r}>0,
\end{equation}
can be rewritten using the Maxwell relations as
\footnote{Note typos in \citet{lattimer_81}, where $\rho^{-2}$
  appears instead of $\rho^{2}$  in their Equations~(5) and (7).}
\begin{equation}
  \left( \frac{\pd P}{\pd s} \right)_{Y_e,\rho}
  \frac{\ud s}{\ud r}
  +\rho^2 \left(\frac{\pd \mu_\nu}{\pd \rho}\right)_{s,Y_e}
  \frac{\ud Y_e}{\ud r}>0.
\end{equation}
Thus, negative lepton number gradients become stabilising if
\begin{equation}
  \left(\frac{\pd \mu_\nu}{\pd \rho}\right)_{s,Y_e}
  =
  \frac{\ud \mu_\nu}{\ud \rho}
  -\left(\frac{\pd \mu_\nu}{\pd s}\right)_{\rho,Y_e}\frac{\ud s}{\ud \rho}
  -\left(\frac{\pd \mu_\nu}{\pd Y_e}\right)_{s,\rho}\frac{\ud Y_e}{\ud \rho}
  <0.
\end{equation}
Since $s \approx \mathrm{const}.$ and
$\ud Y_e/\ud \rho>0$ in the PNS convection zone,
and since $(\pd \mu_\nu/\pd Y_e)_{s,\rho}$ is generally positive,
the $\mu_\nu$-profile indicates that the negative $Y_e$-gradient
is stabilising throughout large parts of the convective layer.

Could the stratification of $Y_e$ and the stabilising role of the
$Y_e$-gradient be related to LESA and the peculiar spectra of the
convective fluctuations?  Double-diffusive instabilities
\citep{bruenn_96,bruenn_04} could in principle lead to considerably
different dynamics compared to the familiar Schwarzschild-Ledoux
convection. However, since LESA also appears in codes like ours that
do not include lateral diffusion and hence do not account for
equilibration of rising and sinking fluid elements with their
surroundings, double-diffusive instabilities cannot be at the heart of
LESA, although they may modify its appearance in detail.

Based on our findings, we speculate that there may nonetheless be a
connection to the stabilising role of negative $Y_e$-gradients, which
may affect convective eddies of various sizes differently: Since the
$Y_e$-gradient is relatively shallow in a narrow region within the
convective zone, small-scale overturn motions in this region would not
be strongly affected by stabilising gradients, whereas turnover
motions spanning the entire convective region will undergo reduced
acceleration by buoyancy forces since the $Y_e$-contrast to their
surroundings partly negates the higher/lower entropy of rising/sinking
bubbles. This might explain the unusually shallow slope
of the turbulent velocity spectrum.

Since the $Y_e$-gradient in this central layer of the PNS convection
zone is shallow, these small-scale overturn motions will be
characterized by small $Y_e$-contrast, and hence contribute little
power to the spectrum of $Y_e$ perturbations on small scales.  Thus,
unlike the spectrum of velocity perturbations, the spectrum
$\hat{Y}_{e,\ell}$ will still exhibit a significant slope.

This explanation is only a tentative one, and we will
address the structure of PNS convection in more detail
in future work. At present our findings mainly serve
to highlight the complexity of the LESA phenomenon beyond
what has already been described in previous studies.
The peculiar features of LESA are also of considerable
importance in the context of gravitational wave emission;
the power in $\ell=2$ modes that can be seen in gravitational
waves is evidently tightly linked to the shape of the turbulence spectra.
LESA excites the g-mode oscillations of the PNS that
subsequently lead to the high frequency gravitational wave emissions.
Further analysis of LESA is thus clearly called for. 

  It is also worth noting that LESA persists until the end of the
simulation in both models (Figure~\ref{fig:lesa_final}), i.e., well
into the explosion phase. Since the accretion rate onto the
PNS has already dropped significantly by that time, this could
be further evidence that the LESA is a phenomenon that arises
within the PNS convection itself, rather than depending on
a feedback process between the accretion flow and PNS convection
as originally envisaged by \citet{tamborra_14}.

\section{Gravitational Wave Emission}
\label{sec:gw}

\subsection{Model He3.5}

\begin{figure*}
\centering
\includegraphics[width=\textwidth]{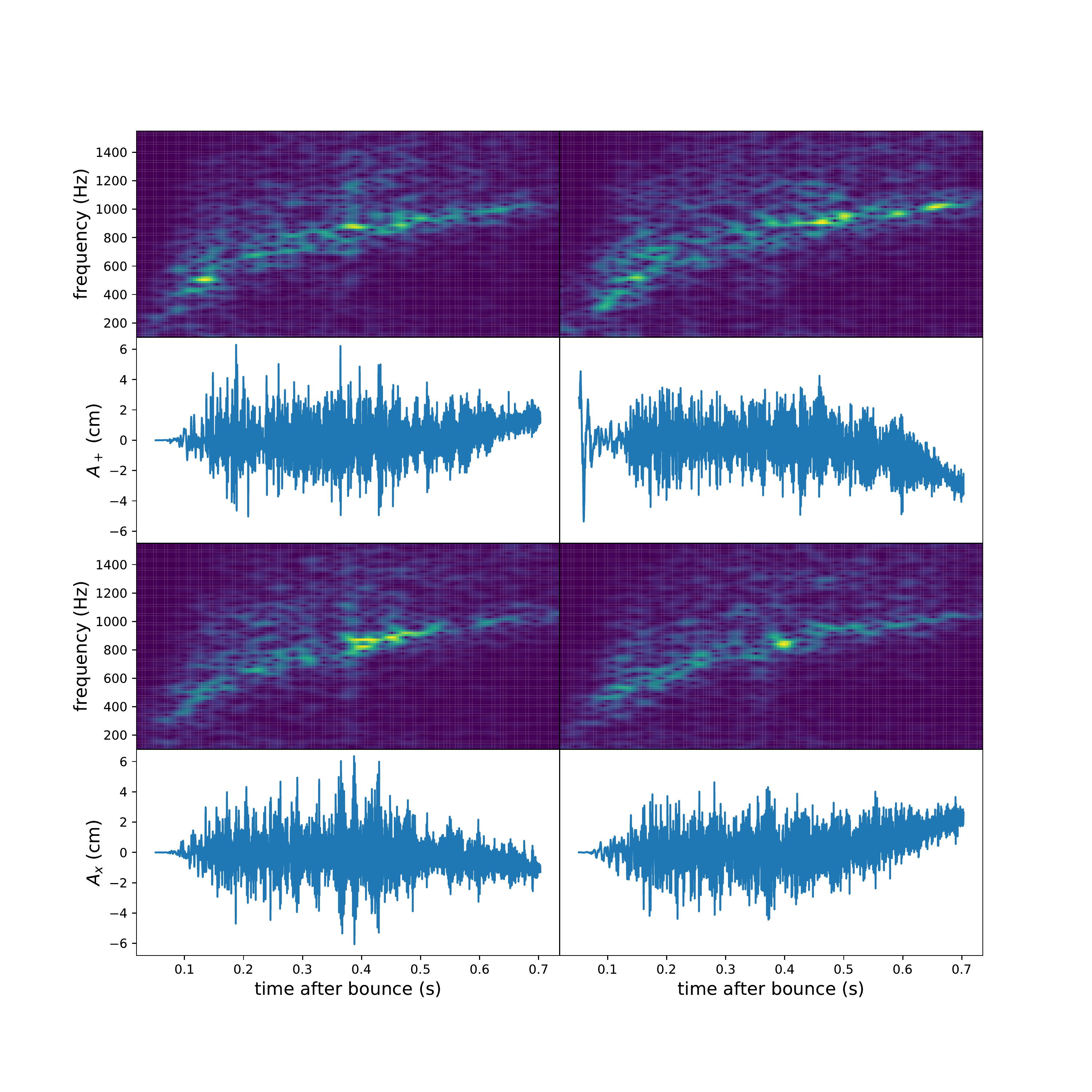}
\caption{The gravitational wave emission for model He3.5. The left column shows the two polarizations at the pole, and the right column shows the two polarizations calculated at the equator. The frequency peaks between 800\,-1000\,Hz. The gravitational wave emission peaks after 0.3\,s during the revival of the shock. A small tail occurs after 0.5\,s.}
\label{fig:s3.5signal}
\end{figure*}

The gravitational wave emission of model He3.5 is shown in Figure \ref{fig:s3.5signal}. We show the time series and spectrograms for both gravitational wave polarizations for observers at the pole $(\theta,\phi) = (0,0)$ and equator. As expected for a non-rotating simulation, there is a quiescent phase up to $\sim0.1$\,s after core-bounce. At this early time there is a spike in the time series of the plus polarization for an observer at the equator. This occurs because the model is adjusting after we start a 3D treatment of the PNS convection zone after switching from 1D. This is followed by a stochastic phase in the time series. The largest gravitational wave amplitudes, of up to 6\,cm, occur after 0.35\,s post bounce as the shock is revived and the model enters the explosion phase. The total energy radiated in gravitational waves is $E_{gw} = 9.0\times 10^{-10} M_{\odot}c^{2}$ for this model, as given by Equation~(22) in \citet{1997A&A...317..140M}.

The signal spectrograms show the gravitational wave emission increases in frequency with time as is consistent with the emission expected from g-mode oscillations of the PNS surface. The peak frequency is higher than other recent models at 800\,-1000\,Hz. Our model has no low frequency emission due to the SASI. This is due to the small core mass and the rapid drop in the accretion rate, similar to the $11.2\,M_{\odot}$ model in \citet{2017MNRAS.468.2032A}. 

A tail in the time series occurs after 0.5\,s. The tail is due to anisotropic expansion of the shock wave. A positive amplitude indicates a prolate explosion, and a negative amplitude indicates an oblate explosion \citep{2009ApJ...707.1173M}. Our models show a more pronounced tail signal than any of the other recent 3D models using multi-group neutrino transport \citep{2016ApJ...829L..14K, 2017MNRAS.468.2032A, 2017arXiv170107325Y, 2018arXiv181007638A}, as these models were not evolved sufficiently far into the explosion phase.

\subsection{Model s18}

\begin{figure*}
\centering
\includegraphics[width=\textwidth]{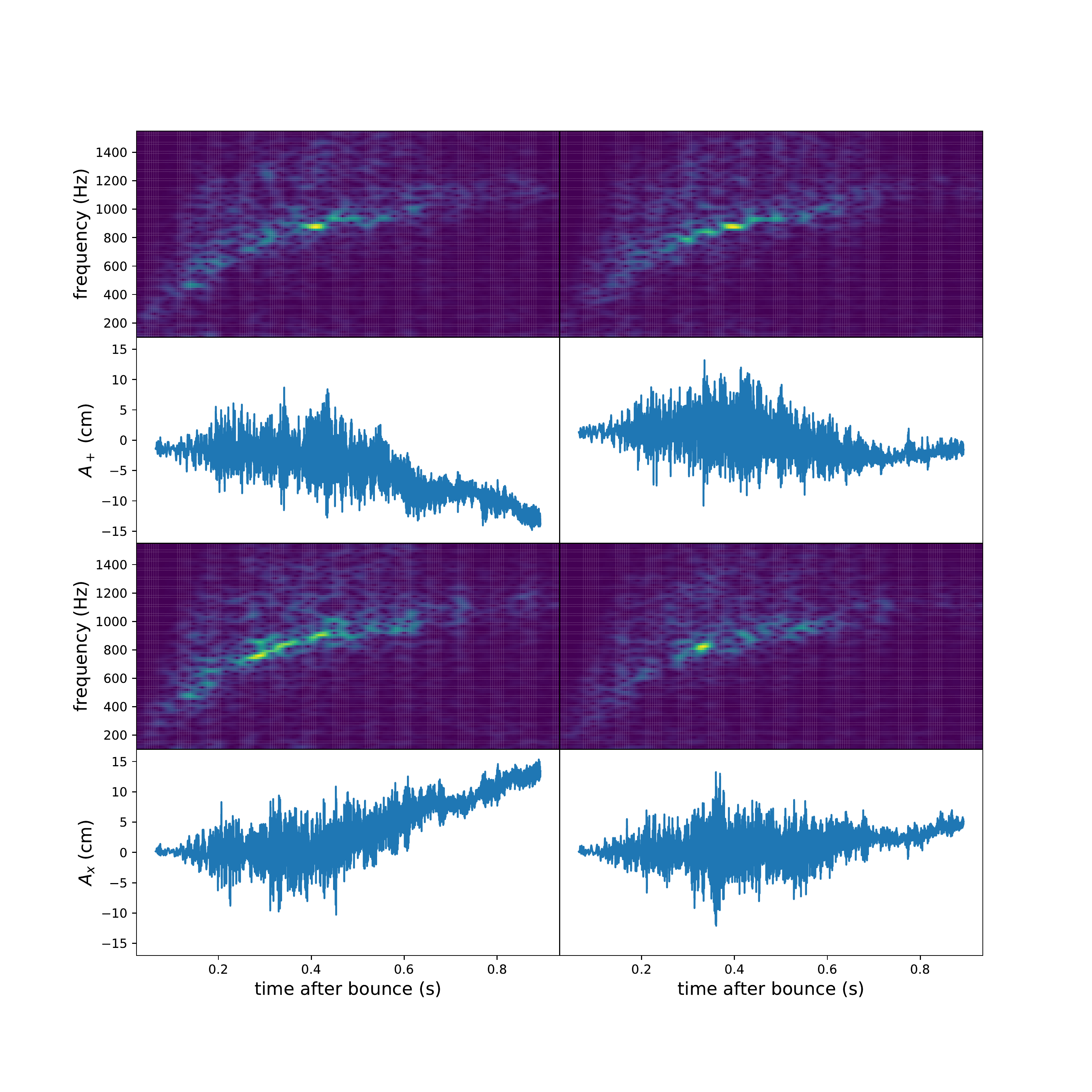}
\caption{The gravitational wave signal for model s18. As for model He3.5, the left column is the gravitational wave emission at the pole, and the right column is the gravitational wave emission at the equator. The peak gravitational wave emission occurs earlier than for model He3.5 due to the earlier shock revival and has similar peak frequencies between 800\,-1000\,Hz. This model has a large low frequency tail at late times.}
\label{fig:s18signal}
\end{figure*}

\begin{figure}
\centering
\includegraphics[width=\linewidth]{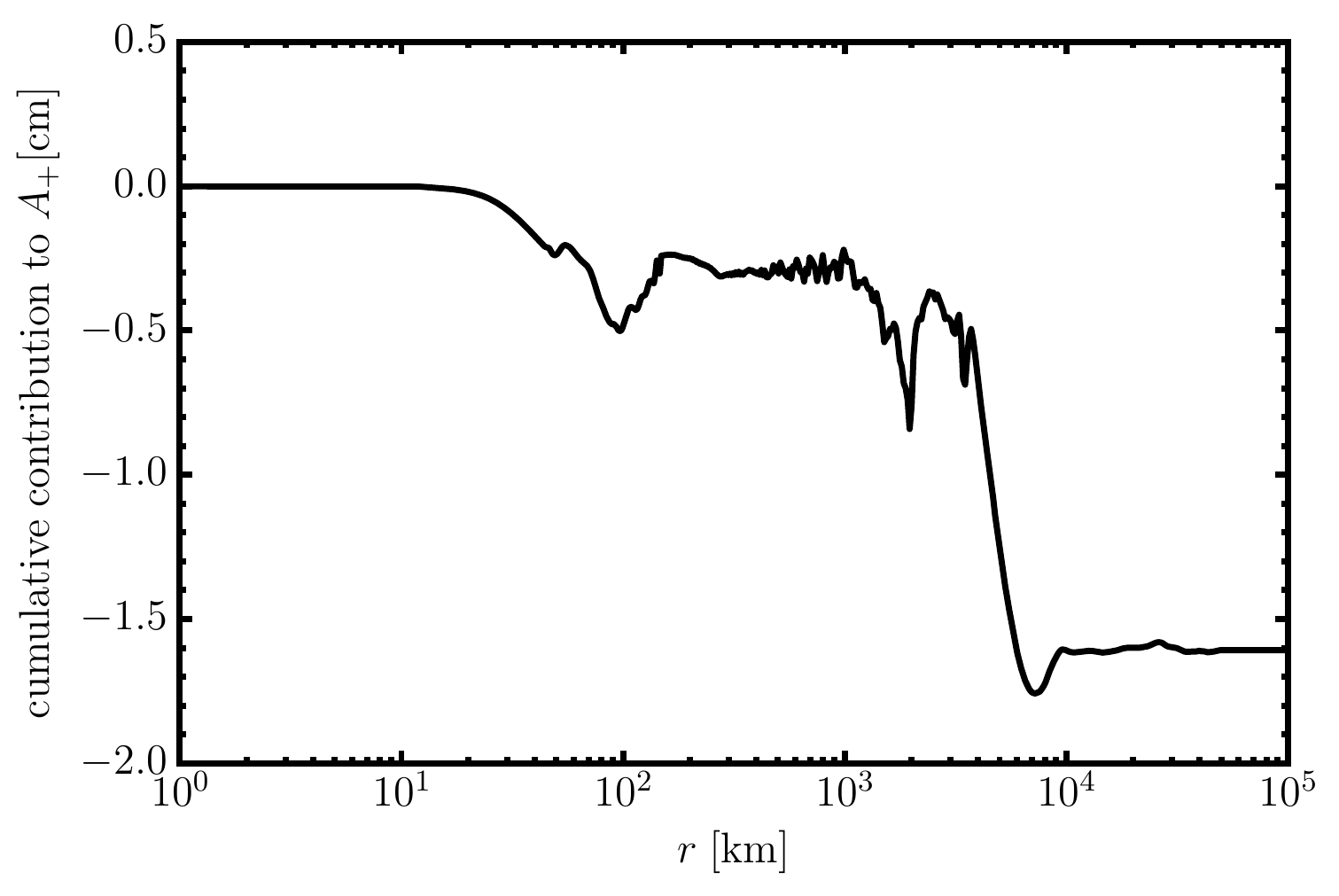}
\caption{Cumulative contribution of the matter within
  radius $r$ to the initial offset of the gravitational wave amplitude $A_+$
  in the polar direction for model s18. The offset
  is mostly due to the quadrupolar deformation of
  the collapsing oxygen shell.
}
\label{fig:offset}
\end{figure}

The gravitational wave emission produced by model s18 is shown in Figure \ref{fig:s18signal}. The model has almost no gravitational wave emission before 0.1\,s after core bounce, except for a small negative offset in the plus polarization at the pole. This offset is physical and is due to the fact that the collapsing
  shells are not spherically symmetric, but inherit a non-negligible quadrupolar deformation from convective burning in the progenitor.
  By evaluating the cumulative contribution of the region inside radius $r$ in the quadrupole formula
  (Figure~\ref{fig:offset}), one finds that the offset is mostly due to the quadrupolar deformation from within the
  violently convective oxygen shell \citep{2016ApJ...833..124M} between  between $r=3000\, \mathrm{km}$ and $r=8000\, \mathrm{km}$, although the shells further inside also become mildly aspherical during the infall and contribute a bit to the offset.
The peak gravitational wave emission starts around 0.3\,s after core bounce after the revival of the shock. The highest amplitudes continue up to 0.5\,s well into the explosion phase. The peak frequency is similar to the previous model, between 800\,-1000\,Hz, but occurs earlier than model He3.5 due to the earlier explosion time. This model has a larger gravitational wave amplitude than most other recent 3D CCSN simulations with a maximum amplitude of $\sim10$\,cm due to the large mass and high explosion energy of this model of $\sim0.8$\,Bethe. This translates to a gravitational wave energy of $E_{gw} = 3.3\times 10^{-9} M_{\odot}c^{2}$. 

The gravitational wave emission is due to oscillations of g-modes in the PNS. There is no low frequency emission due to the SASI because of the strong density and pressure perturbations ahead of the shock from convective burning in the oxygen shell. These perturbations disrupt the amplification cycle of vorticity waves and acoustic waves that normally lead to gravitational wave emission produced by the effects of the SASI. This is discussed in more detail in the previous simulation of this model in \citet{2017MNRAS.472..491M}. There is a large tail in the time series which occurs after 0.45\,s. 

\subsection{The g-mode excitation mechanism and GW strength}

\begin{figure}
\centering
\includegraphics[width=\columnwidth]{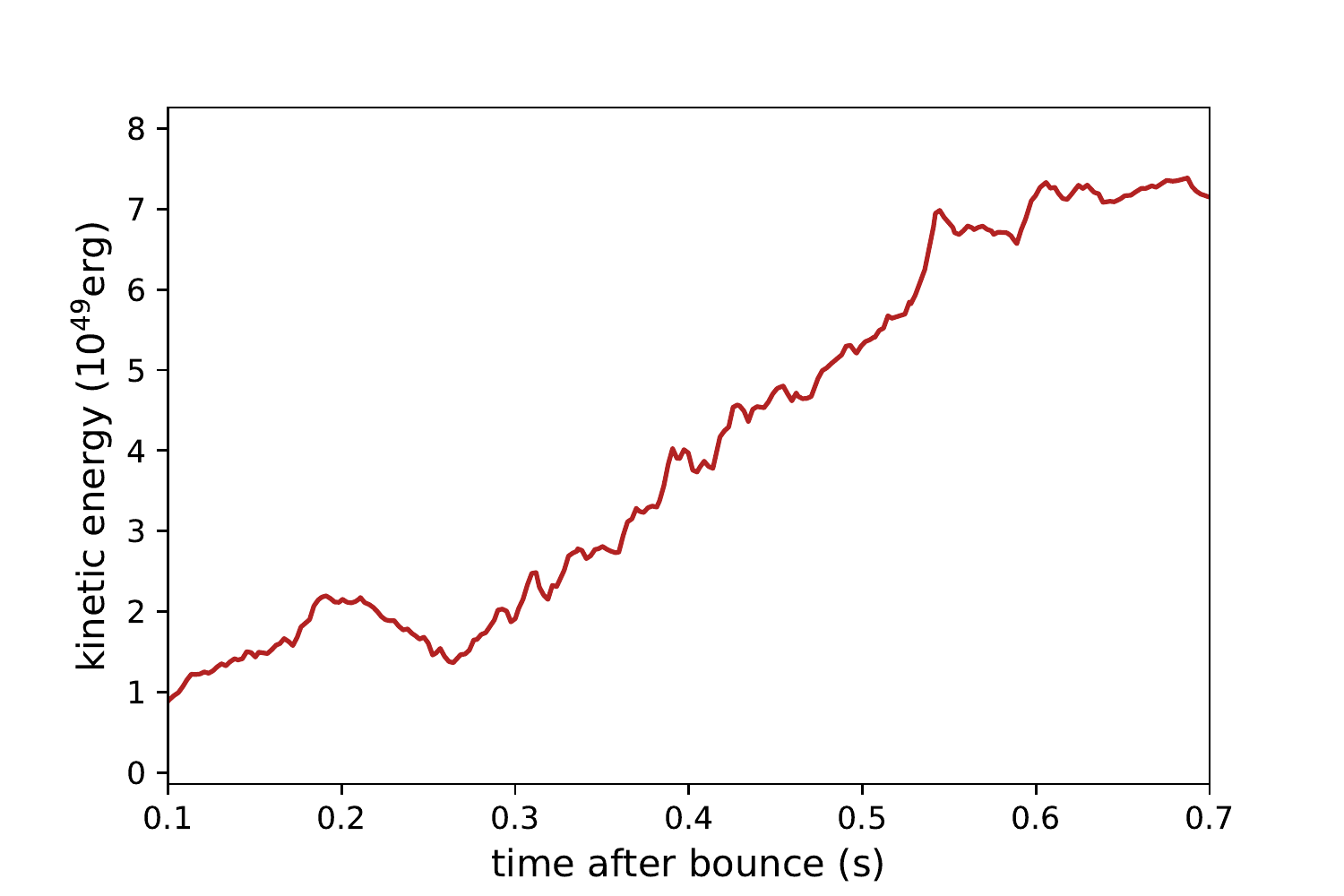}
\caption{The time evolution of the non-radial kinetic energy
contributed from velocity fluctuations in the PNS
convection zone for model s18. PNS convection is still 
energetic after the shock revival phase. }
\label{fig:kinetic_energy}
\end{figure}

The fact that the high-frequency gravitational wave emission peaks shortly after
  shock revival in both models and then declines visibly is relevant
  for the ongoing discussion about the excitation mechanism of the PNS
  surface g-mode.  \cite{2017MNRAS.468.2032A} determined that g-modes
  are predominantly excited by PNS convection, but the rise and
  decline of the high-frequency emission in our models rather points
  to a correlation of the emission with the violence of non-radial
  fluid motions in the gain region, somewhat more similar to 2D
  explosion models.  This is reinforced by the finding that the
  kinetic energy in non-radial motions inside the PNS convection zone
  actually \emph{grows} continuously in model s18
  (Figure~\ref{fig:kinetic_energy}) even while the high-frequency
  emission is already subsiding. Recently, \citet{radice_18} also
  observed such a correlation and found that there is even a
  relatively tight quantitative relationship between the total energy
  $E_\mathrm{GW}$ of emitted gravitational waves and the time-integrated flux of
  turbulent kinetic energy $E_\mathrm{turb}$ from the gain region onto
  the PNS,
  \begin{equation}
\label{eq:scaling}
    E_\mathrm{GW}\approx 3\times 10^{43} \left( \frac{E_\mathrm{turb}}{10^{50}\, \mathrm{erg}}\right)^{1.88}.
  \end{equation}

The disparate findings on the relative contribution of PNS convection
and non-radial motions in the gain layer are not at odds with each other, however.
It is natural to expect that the relative importance of these two possible
drivers of surface g-modes varies across models depending on the violence
of non-radial motions in the different unstable regions. One can
also derive this more quantitatively \citep{mueller_17} by
estimating the energy in g-modes using analytic theory,
which suggests that the g-mode energy flux $L_g$ scales
with the convective luminosity $L_\mathrm{conv}$ in adjacent convective
regions and the convective Mach number $\mathrm{Ma}$ as \citep{goldreich_90}
\begin{equation}
L_g\sim \mathrm{Ma}\, L_\mathrm{conv}.
\end{equation}
By assuming that the forcing of the g-modes is coherent
over one convective turnover time-scale $\tau$,
we can express the energy $E_g$ in g-modes in terms
of the convective energy  $E_\mathrm{conv}$ in the
forcing motions as \citep{mueller_17}
\begin{equation}
\label{eq:eg}
E_g \sim  \mathrm{Ma}\, L_\mathrm{conv}  \tau
\sim \mathrm{Ma}\, (E_\mathrm{conv}/\tau)  \tau
\sim \mathrm{Ma}\, E_\mathrm{conv}.
\end{equation}
The convective luminosity in the PNS convection zone
is of the order of the diffusive neutrino luminosity $L_\mathrm{core}$
from the PNS core, which does not vary tremendously across time
and progenitors \citep{mirizzi_16}. 
The convective luminosity in the gain region, however,
scales with the neutrino heating rate $\dot{Q}_\nu$
 \citep{murphy_12,mueller_15a},
\begin{equation}
L_\mathrm{conv} \sim \dot{Q}_\nu,
\end{equation}
which varies considerably across progenitors and between
the pre-explosion and explosion phase. Since
$\mathrm{Ma}\, L_\mathrm{conv}  \tau$
is also of the same order in the PNS convection zone
and the gain region (where $L_\mathrm{conv}$ is lower
by a factor of $\mathcal{O}(10)$, but $\mathrm{Ma}$
and $\tau$ are higher by a factor of several), the relative
importance of g-mode excitation by PNS convection and
non-radial motions in the gain region is bound to
vary quite significantly.  If $\mathrm{Ma}$ and $\dot{Q}_\nu$ are high
-- as in our more energetic exploding models 
and those of \citet{radice_18} -- excitation of
g-modes by turbulent motions in the gain region is more likely
to dominate than in the pre-explosion phase, which was the primary
focus of \citet{2017MNRAS.468.2032A}.

Even though Equation~(\ref{eq:eg}) only provides a very crude model
for the interaction between convection and/or SASI and the surface
g-mode, it seems to provide a very natural way of interpreting
the results on gravitational wave emission from the various recent 3D models
(especially \citealt{2017MNRAS.468.2032A}, \citealt{radice_18}, and this paper)
and naturally explains the scaling relation
(\ref{eq:scaling}) found by \citet{radice_18}: Using dimensional analysis,
one can determine that the typical gravitational wave amplitudes scale with $E_g$
as \citep{mueller_17}
\begin{equation}
A\sim\frac{\alpha GE_g}{c^4},
\end{equation}
where the dimensionless factor $\alpha$ is of order unity
and accounts for the fact that only part of the g-mode energy
is stored in quadrupolar modes. If $f$ is the typical emission frequency, the gravitational wave luminosity
$\dot{E}_\mathrm{GW} $
is roughly 
\begin{equation}
\dot{E}_\mathrm{GW} \sim
\frac{4\pi c^3 f^2 A^2}{G}.
\end{equation}
After integrating over a typical time $T$ of strong emission and inserting
Equation~(\ref{eq:eg}) for $L_\mathrm{conv}$, we obtain
$E_\mathrm{GW}$ in terms of the time-integrated flux
$E_\mathrm{turb}=L_\mathrm{conv} T$
of turbulent kinetic energy onto the PNS surface
as
\begin{multline}
E_\mathrm{GW} \sim
\nonumber
\frac{4\pi G (f \tau \mathrm{Ma} L_\mathrm{conv} T)^2}{c^5 }
=
\frac{4\pi G f^2 \tau^2 \mathrm{Ma}^2 E_\mathrm{turb}^2}{c^5 }
\\
=
4.2 \times 10^{43} \, \mathrm{erg}
\left(\frac{f}{1000\, \mathrm{Hz}}\right)^2
\left(\frac{\tau}{20\, \mathrm{ms}}\right)^2
\left(\frac{\mathrm{Ma^2}}{0.3}\right)
\left(\frac{E_\mathrm{turb}}{10^{50}\,\mathrm{erg}}\right)^2.
\end{multline}
For typical values around shock revival of $\tau=20\, \mathrm{ms}$
\citep{mueller_14} and $\mathrm{Ma}^2=0.3$
\citep{mueller_15a}, the result is very similar to the empirical
scaling relationship of \citet{radice_18} with only a slightly
different power-law exponent.

\section{Detection Prospects}
\label{sec:detection}

\begin{figure}
\centering
\includegraphics[width=\columnwidth]{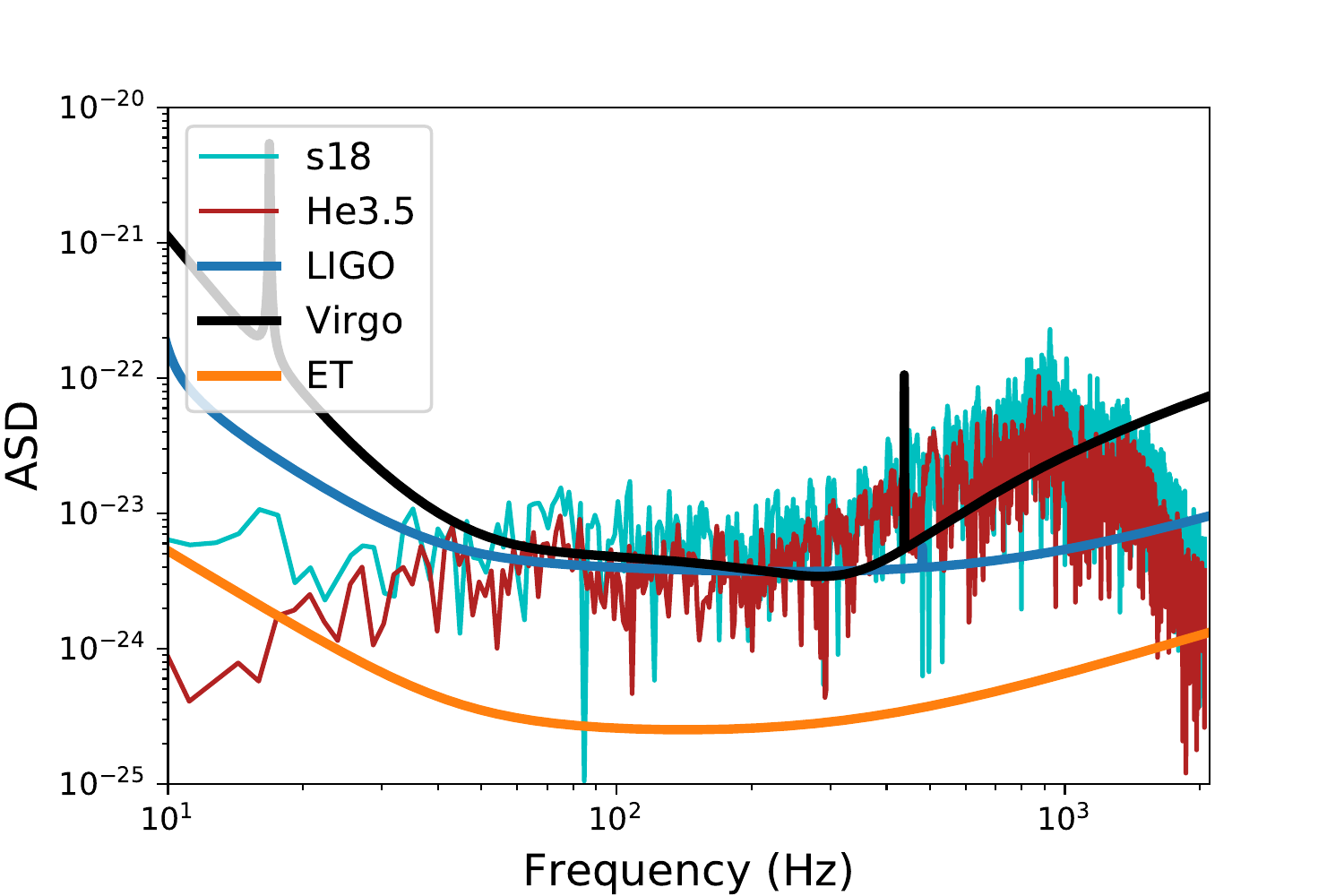}
\caption{The gravitational wave models 
at a distance of 10\,kpc relative to the design sensitivity
amplitude spectral density (ASD) curves of 
Advanced LIGO, Advanced Virgo, and ET-B, assuming maximum antenna pattern 
sensitivity of the signals. ET will improve the detection 
sensitivity across the entire frequency band. }
\label{fig:noisecurve}
\end{figure}

In this section, we estimate the distance out to which our gravitational wave signals may be detected in current ground-based gravitational wave detectors and a planned next generation underground detector called Einstein Telescope (ET; \citealp{0264-9381-27-19-194002}). We simulate Gaussian design sensitivity noise for the aLIGO and AdVirgo detectors using the power spectral density (PSD) curves described in \citet{2018LRR....21....3A}. For ET, we use the ET-B noise curve described in \citet{2008arXiv0810.0604H}. The noise curves are shown in Figure \ref{fig:noisecurve}, as well as the maximum amplitude spectral density (ASD) of our models at a distance of 10\,kpc. The AdVirgo detector has poor sensitivity above 400\,Hz, in the region where our models have the greatest gravitational wave amplitude. The aLIGO detectors have better sensitivity at the higher peak frequencies in our models. In the case of ET, the sensitivity improves over the entire frequency band of our models greatly improving the detection prospects.  

To make an approximate estimation of the distance out to which our waveforms can be detected, we use the Bayesian inference code known as Bilby \citep{bilby}. Bilby is most commonly used as a parameter estimation tool, however the Bayes factors calculated in Bilby can be used as a detection statistic \citep{2018PhRvX...8b1019S, PhysRevD.95.104046}. We use a sine Gaussian signal model, as this is the signal model typically used in searches for gravitational wave burst signals in aLIGO and AdVirgo data \citep{2016PhRvD..94j2001A, PhysRevD.95.104046}. The sine Gaussian signal model is defined in the frequency domain as
\begin{equation}
h_{+} (f) = \frac{h_{rss}}{\sqrt{ \frac{Q}{4\sqrt{\pi}f_{o}} (1+e^{-Q^{2}}) }} \frac{\sqrt{\pi} \tau}{2} (e^{-(f-f_{0})^{2}\pi^{2}\tau^{2}}+e^{-(f+f_{0})^{2}\pi^{2}\tau^{2}})
\end{equation}
\begin{equation}
h_{\times} (f) = -i\frac{h_{rss}}{\sqrt{ \frac{Q}{4\sqrt{\pi}f_{o}} (1-e^{-Q^{2}}) }} \frac{\sqrt{\pi} \tau}{2} (e^{-(f-f_{0})^{2}\pi^{2}\tau^{2}}-e^{-(f+f_{0})^{2}\pi^{2}\tau^{2}})
\end{equation}
where $Q$ is the quality factor, $\tau$ is the duration, $f$ is an array of frequencies, and $f_{0}$ is the central frequency. The root-sum-squared amplitude $h_\mathrm{rss}$ is given by the sum of the amplitude of the individual polarizations multiplied by their corresponding detector antenna pattern values which are dependent on the sources sky position and GPS time.  
 
We assume that the signal start time is known within 3\,s due to 
a coincident neutrino detection. We ``inject'' the signals into the data as measured at the pole assuming a sky location of the galactic center. Multiple GPS times are used to cover a range of detector sensitivities, as the sensitivity of the gravitational wave detectors varies at different times and sky positions due to their antenna patterns. We then vary the source distance and calculate Bayes factors that determine the probability of the data containing a signal plus noise or containing only noise. The evidence $Z_{M}$ for a model $M$, with parameters $\theta$, in data $d$ is given by
\begin{equation}
Z_{M} = p(d|M,I) = \int p(d|\theta,M,I)p(\theta|M,I)d\theta
\end{equation}
where $p(d|\theta,M,I)$ is the likelihood and $p(\theta|M,I)$ is the prior. As in real gravitational wave burst searches, we use a Gaussian likelihood function and uniform priors for our sine Gaussian signal parameters $Q$ and $f_{0}$, a uniform on the sky prior for the sky position, and uniform in volume prior for the signal amplitude. Bilby then calls a nested sampler to solve the integral. This transforms the integral into
\begin{equation} 
Z_{M} = \sum^{N}_{i} p(d|\theta,M,I) w_{i},
\end{equation}
where $w$ is the fraction of the prior occupied by point i. An odds ratio that is equvilant to a Bayes factor can then be calculated as 
\begin{equation}
\mathcal{O} = \frac{Z_{S}}{Z_{N}} = \frac{p(d|M_{S},I)}{p(d|M_{N},I)}  \frac{p(M_{S},I)}{p(M_{N},I)},
\end{equation}
where $M_{S}$ and $Z_{S}$ are the signal model and evidence, and $M_{N}$ and $Z_{N}$ are the noise model and evidence. We consider a signal as being detected if we obtain a log Bayes factor larger then 8. 

\begin{figure*}
\centering
\includegraphics[width=\columnwidth]{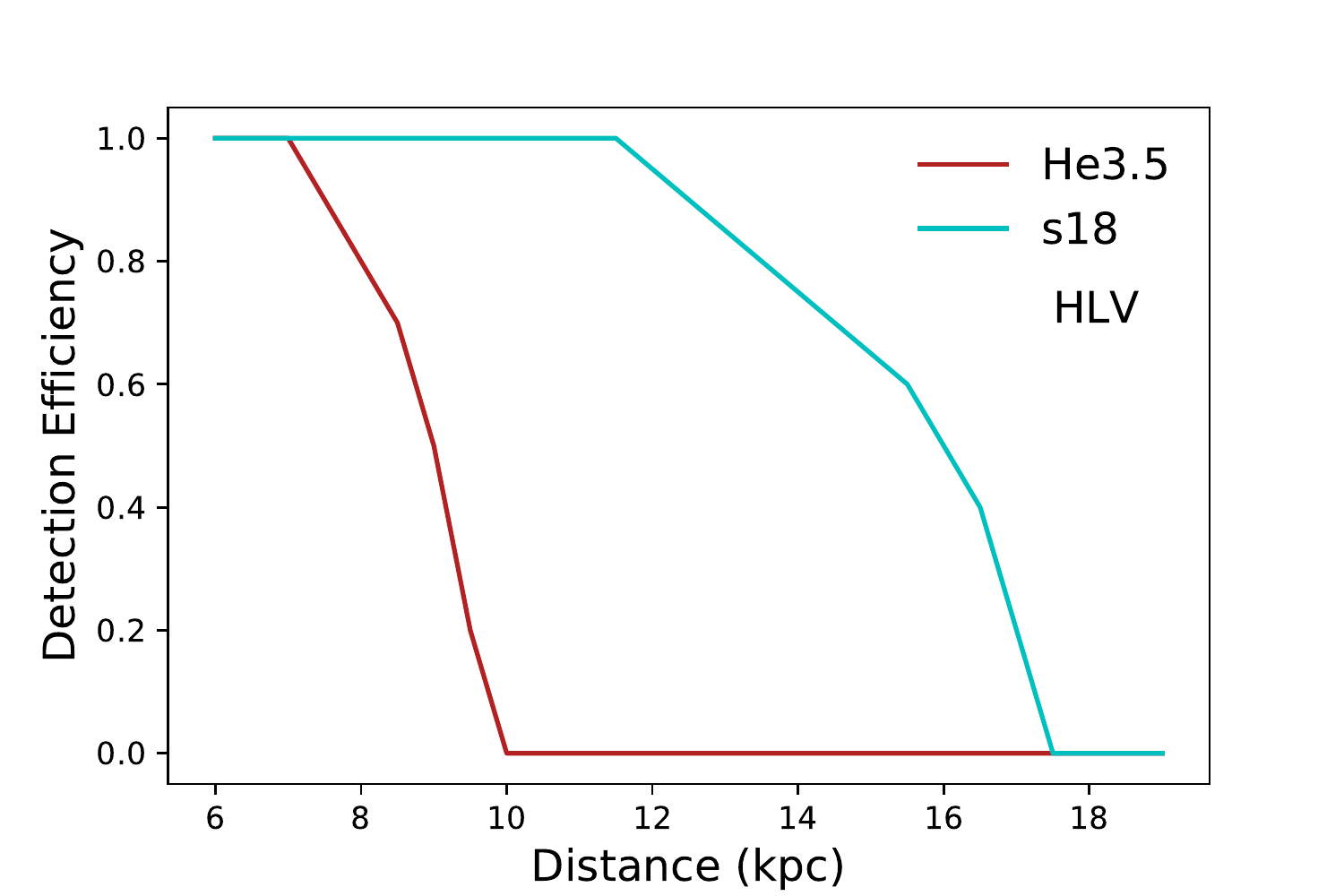}
\includegraphics[width=\columnwidth]{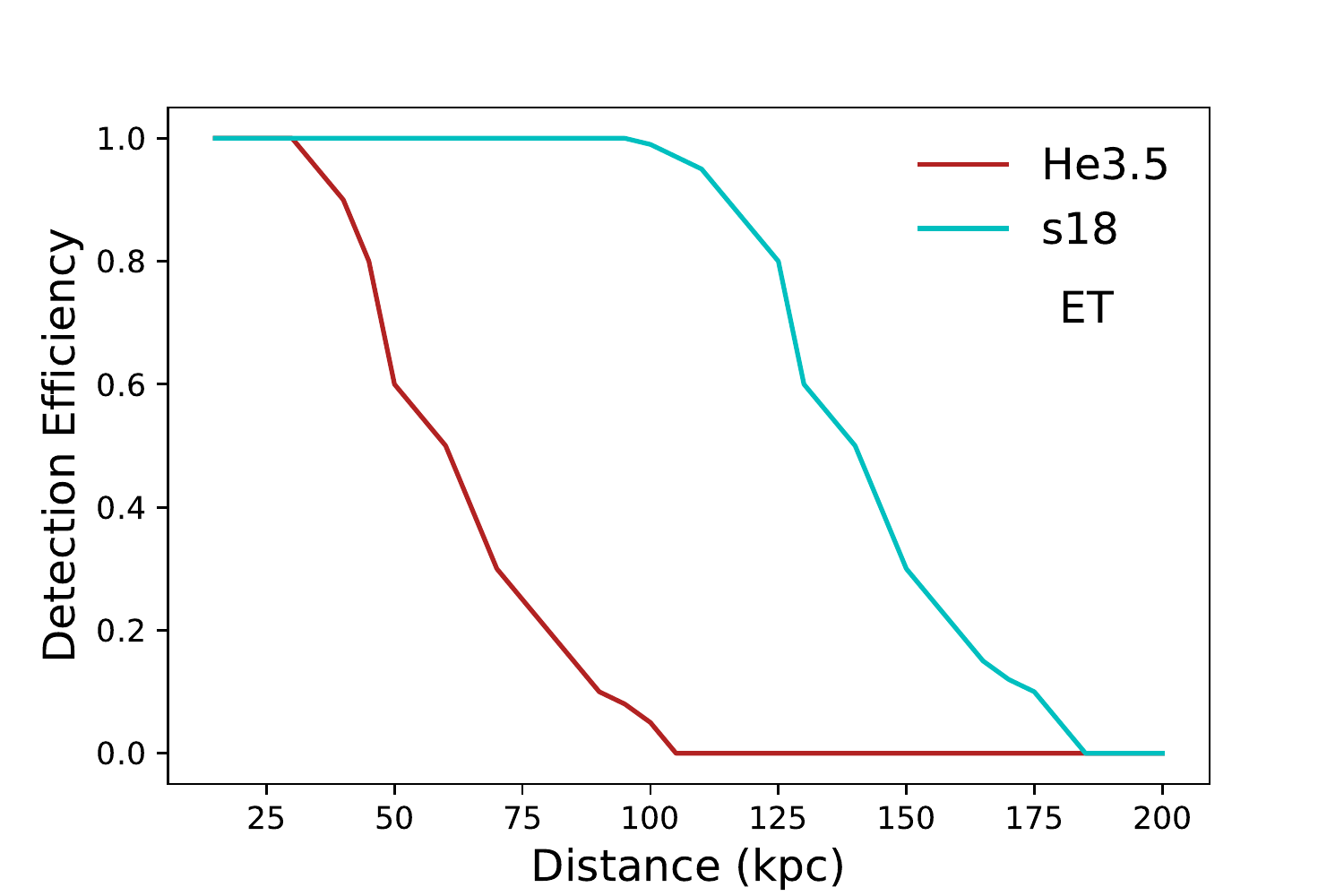}
\caption{An estimation of the detection distance of models s18 and He3.5 in an Advanced LIGO/Virgo network (left) and the Einstein Telescope detector (right). The maximum detectable distance for model He3.5 is 10\,kpc in an Advanced LIGO/Virgo network, and 105\,kpc in Einstein Telescope. The maximum detectable distance for model s18 is 17.5\,kpc in an Advanced LIGO/Virgo network, and 180\,kpc in Einstein Telescope.  }
\label{fig:detections}
\end{figure*}

The results are shown in Figure \ref{fig:detections}. In an aLIGO and AdVirgo design sensitivity detector network, we estimate the maximum detectable distance to be 10\,kpc for model He3.5, and 17.5\,kpc for model s18. The maximum detectable distance in ET data is 105\,kpc for model He3.5, and 180\,kpc for model s18. This shows that our models are detectable in our Galaxy and beyond the Large Magellanic Cloud with the ET.  

We find that our models have larger detection distances than the results in previous detectability studies, which include emission from older neutrino-driven simulations \citep{2016PhRvD..93d2002G, 2016PhRvD..94j2001A}. CCSN 2D models result in artificially weak gravitational wave energy in the detector due to the single polarization, leading to smaller detection distances. Our new models can be detected at larger distances than the models in \citet{2013ApJ...768..115O} due to our models higher amplitudes and longer durations. The models in \citet{2012A&A...537A..63M} are also of long duration.
However, while this model initially also develops strong large-scale
convective motions and has similarly strong shock deformation
as ours (as exemplified by a tail of similar amplitude),
the detection distance of our models are larger. This is because
our models exhibit stronger large-scale convective overturn for a 
longer time after shock revival, whereas accretion downflows are
quenched more rapidly in the model of \citet{2012A&A...537A..63M}.
Previous simulated models with lower peak gravitational wave emission produced by the low-frequency $(\sim100\,\mathrm{Hz})$ oscillatory SASI motions in the post-shock regions and the PNS surface are likely more detectable in AdVirgo than our models that peak in a higher frequency band where AdVirgo does not have much sensitivity. However, the aLIGO detectors have better sensitivity in the peak frequency range of our models, and ET has better sensitivity across the whole frequency band of all current neutrino-driven CCSN models.

\section{Conclusions}
\label{sec:conclusion}

In this study, we simulate the explosion of an ultra-stripped star with a helium star mass of $3.5\,M_{\odot}$ and a single star with a ZAMS mass of $18\,M_{\odot}$. We update the neutrino hydrodynamics code \textsc{CoCoNuT-FMT} to include updated neutrino rates and on-the-fly extraction of the gravitational wave emission. We simulate in 3D down to the innermost 10\,km to include the PNS convection zone, as this region is essential for accurate gravitational wave emission predictions. To keep the computational costs manageable, we use a reduced number of energy groups than in recent \textsc{CoCoNuT-FMT} models. 

Model He3.5 is the first ultra-stripped progenitor simulation to include a calculation of the gravitational wave emission. Understanding ultra-stripped CCSNe is essential as most stars are expected to occur in binary systems. We find our model explodes later than the previous simulation in \citet{2018arXiv181105483M} due to the lower number of energy groups in our simulation. The model has gravitational wave emission associated with the excitation of g-modes in the PNS with peak frequencies between 800\,Hz and 1000\,Hz. As none of the properties that determine the frequency of the emission were affected by the lower number of energy groups, we determine that our peak frequencies are still accurate to within a few percent. As the peak gravitational wave amplitude, of $\mathord{\sim} 6\,\mathrm{cm}$, occurs after shock revival, our reduced number of energy groups may have resulted in our peak gravitational wave amplitude occurring a little later after core bounce than if a larger number of energy groups were included. 

Model s18 reaches even further into the explosion phase as the simulation was stopped 0.9\,s after core bounce. There are no significant differences between the explosion dynamics of the model and the previous simulation with a larger number of energy groups. The gravitational wave emission is similar to model He3.5. The emission is due to g-modes oscillations of the PNS, and the frequency peaks between 800\,Hz and 1000\,Hz. The peak gravitational wave emission begins earlier for this model due to the earlier explosion time, and the larger mass and higher explosion energy of model s18 results in a larger peak gravitation wave amplitude, up to $\sim 10$\,cm.

Other recent 3D simulations by \citet{2017MNRAS.468.2032A} and \citet{2017arXiv170107325Y} have varied wildly in their estimation of the amplitude of their gravitational wave models. We find the amplitude of our models to be closer to those of \citet{2017MNRAS.468.2032A}, even though our models successfully revive the shock and have greater explosion energy. 

We include a discussion of the LESA instabilities observed in our models. We find evidence that the LESA may not be a new instability, but just a manifestation of regular convection. Our analysis highlights the complexity of the LESA and the need for further analysis in future studies to fully understand the effects of the LESA on the gravitational wave emission.  

We measure how detectable the gravitational wave emission in our models would be in an aLIGO and AdVirgo design sensitivity detector network, and in the next generation detector ET. We find our larger gravitational wave amplitudes result in detection distances of up to 10\,kpc for model He3.5, and up to 17.5\,kpc for model s18 in an advanced detector network. In ET, both models are detectable at distances beyond the Large Magellanic Cloud. The detectable distance for our models may improve in the future if more sensitive search techniques for CCSN signals are developed that are sensitive to lower signal-to-noise ratio signals. 

Our simulations stop after the peak gravitational wave emission phase and are longer in duration than all of the recent neutrino-driven simulations produced in \citet{2018arXiv181007638A, 2016ApJ...829L..14K, 2017arXiv170107325Y, 2017MNRAS.468.2032A}. This is important for several reasons. The first is that knowledge of the entire gravitational waveform is essential for accurate tuning of gravitational wave searches and parameter estimation algorithms such as those in \citet{2017PhRvD..96l3013P, 2018arXiv180207255G, 2018MNRAS.474.5272T, 2016PhRvD..93d2002G, 2016PhRvD..94j2001A}. The second is that most of the previous 3D waveforms are produced in simulations that either do not successfully revive the shock, or are stopped too early to reach the shock revival phase. Our models show the peak gravitational wave emission occurs after shock revival, therefore some previous neutrino-driven simulations may have ended before their peak gravitational wave emission occurred. The frequency of the emission produced in g-mode oscillations increases with time, which may explain why our models have a higher peak gravitational wave frequency than previous work with shorter end times.

\section*{Acknowledgements}

JP is supported by the Australian Research Council (ARC) Centre of Excellence for Gravitational Wave Discovery (OzGrav), through project number CE170100004. BM is supported by ARC Future Fellowship FT160100035. This work was performed on the OzStar supercomputer at Swinburne University of Technology and the Pawsey Supercomputing Centre with funding from the Australian Government. 


\bibliographystyle{mnras}
\interlinepenalty=10000
\bibliography{bibfile}

\begin{thebibliography}{}
\makeatletter
\relax
\def\mn@urlcharsother{\let\do\@makeother \do\$\do\&\do\#\do\^\do\_\do\%\do\~}
\def\mn@doi{\begingroup\mn@urlcharsother \@ifnextchar [ {\mn@doi@}
  {\mn@doi@[]}}
\def\mn@doi@[#1]#2{\def\@tempa{#1}\ifx\@tempa\@empty \href
  {http://dx.doi.org/#2} {doi:#2}\else \href {http://dx.doi.org/#2} {#1}\fi
  \endgroup}
\def\mn@eprint#1#2{\mn@eprint@#1:#2::\@nil}
\def\mn@eprint@arXiv#1{\href {http://arxiv.org/abs/#1} {{\tt arXiv:#1}}}
\def\mn@eprint@dblp#1{\href {http://dblp.uni-trier.de/rec/bibtex/#1.xml}
  {dblp:#1}}
\def\mn@eprint@#1:#2:#3:#4\@nil{\def\@tempa {#1}\def\@tempb {#2}\def\@tempc
  {#3}\ifx \@tempc \@empty \let \@tempc \@tempb \let \@tempb \@tempa \fi \ifx
  \@tempb \@empty \def\@tempb {arXiv}\fi \@ifundefined
  {mn@eprint@\@tempb}{\@tempb:\@tempc}{\expandafter \expandafter \csname
  mn@eprint@\@tempb\endcsname \expandafter{\@tempc}}}

\bibitem[\protect\citeauthoryear{{Abbott} et~al.,}{{Abbott}
  et~al.}{2016}]{2016PhRvD..94j2001A}
{Abbott} B.~P.,  et~al., 2016, \mn@doi [\prd] {10.1103/PhysRevD.94.102001},
  \href {http://adsabs.harvard.edu/abs/2016PhRvD..94j2001A} {94, 102001}

\bibitem[\protect\citeauthoryear{{Abbott} et~al.,}{{Abbott}
  et~al.}{2018}]{2018LRR....21....3A}
{Abbott} B.~P.,  et~al., 2018, \mn@doi [Living Reviews in Relativity]
  {10.1007/s41114-018-0012-9}, \href
  {http://adsabs.harvard.edu/abs/2018LRR....21....3A} {21, 3}

\bibitem[\protect\citeauthoryear{{Abdikamalov}, {Gossan}, {DeMaio}  \&
  {Ott}}{{Abdikamalov} et~al.}{2014}]{2014PhRvD..90d4001A}
{Abdikamalov} E.,  {Gossan} S.,  {DeMaio} A.~M.,   {Ott} C.~D.,  2014, \mn@doi
  [\prd] {10.1103/PhysRevD.90.044001}, \href
  {http://esoads.eso.org/abs/2014PhRvD..90d4001A} {90, 044001}

\bibitem[\protect\citeauthoryear{{Acernese} \& et al.}{{Acernese} \&
  et~al.}{2015}]{AdVirgo}
{Acernese} F.,  et al. 2015, \mn@doi [Classical and Quantum Gravity]
  {10.1088/0264-9381/32/2/024001}, \href
  {http://adsabs.harvard.edu/abs/2015CQGra..32b4001A} {32, 024001}

\bibitem[\protect\citeauthoryear{{Andresen}, {M{\"u}ller}, {M{\"u}ller}  \&
  {Janka}}{{Andresen} et~al.}{2017}]{2017MNRAS.468.2032A}
{Andresen} H.,  {M{\"u}ller} B.,  {M{\"u}ller} E.,   {Janka} H.-T.,  2017,
  \mn@doi [\mnras] {10.1093/mnras/stx618}, \href
  {http://adsabs.harvard.edu/abs/2017MNRAS.468.2032A} {468, 2032}

\bibitem[\protect\citeauthoryear{{Andresen}, {M{\"u}ller}, {Janka}, {Summa},
  {Gill}  \& {Zanolin}}{{Andresen} et~al.}{2019}]{2018arXiv181007638A}
{Andresen} H.,  {M{\"u}ller} E.,  {Janka} H.~T.,  {Summa} A.,  {Gill} K.,
  {Zanolin} M.,  2019, \mn@doi [\mnras] {10.1093/mnras/stz990}, \href
  {https://ui.adsabs.harvard.edu/abs/2019MNRAS.486.2238A} {486, 2238}

\bibitem[\protect\citeauthoryear{{Ashton} et~al.,}{{Ashton}
  et~al.}{2019}]{bilby}
{Ashton} G.,  et~al., 2019, \apjs, 241

\bibitem[\protect\citeauthoryear{{Beck} et~al.,}{{Beck}
  et~al.}{2012}]{2012Natur.481...55B}
{Beck} P.~G.,  et~al., 2012, \mn@doi [\nat] {10.1038/nature10612}, \href
  {http://cdsads.u-strasbg.fr/abs/2012Natur.481...55B} {481, 55}

\bibitem[\protect\citeauthoryear{{Blondin} \& {Mezzacappa}}{{Blondin} \&
  {Mezzacappa}}{2006}]{2006ApJ...642..401B}
{Blondin} J.~M.,  {Mezzacappa} A.,  2006, \mn@doi [\apj] {10.1086/500817},
  \href {http://adsabs.harvard.edu/abs/2006ApJ...642..401B} {642, 401}

\bibitem[\protect\citeauthoryear{Blondin, Mezzacappa  \& DeMarino}{Blondin
  et~al.}{2003}]{0004-637X-584-2-971}
Blondin J.~M.,  Mezzacappa A.,   DeMarino C.,  2003, The Astrophysical Journal,
  584, 971

\bibitem[\protect\citeauthoryear{{Bollig}, {Janka}, {Lohs},
  {Mart{\'{\i}}nez-Pinedo}, {Horowitz}  \& {Melson}}{{Bollig}
  et~al.}{2017}]{bollig_17}
{Bollig} R.,  {Janka} H.-T.,  {Lohs} A.,  {Mart{\'{\i}}nez-Pinedo} G.,
  {Horowitz} C.~J.,   {Melson} T.,  2017, \mn@doi [Physical Review Letters]
  {10.1103/PhysRevLett.119.242702}, \href
  {http://adsabs.harvard.edu/abs/2017PhRvL.119x2702B} {119, 242702}

\bibitem[\protect\citeauthoryear{{Bruenn} \& {Dineva}}{{Bruenn} \&
  {Dineva}}{1996}]{bruenn_96}
{Bruenn} S.~W.,  {Dineva} T.,  1996, \mn@doi [\apjl] {10.1086/309921}, \href
  {http://adsabs.harvard.edu/abs/1996ApJ...458L..71B} {458, L71}

\bibitem[\protect\citeauthoryear{{Bruenn}, {Raley}  \& {Mezzacappa}}{{Bruenn}
  et~al.}{2004}]{bruenn_04}
{Bruenn} S.~W.,  {Raley} E.~A.,   {Mezzacappa} A.,  2004, ArXiv Astrophysics
  e-prints, \href {http://adsabs.harvard.edu/abs/2004astro.ph..4099B} {}

\bibitem[\protect\citeauthoryear{{Buras}, {Rampp}, {Janka}  \&
  {Kifonidis}}{{Buras} et~al.}{2006a}]{buras_06a}
{Buras} R.,  {Rampp} M.,  {Janka} H.-T.,   {Kifonidis} K.,  2006a, \mn@doi
  [\aap] {10.1051/0004-6361:20053783}, \href
  {http://adsabs.harvard.edu/abs/2006A%26A...447.1049B} {447, 1049}

\bibitem[\protect\citeauthoryear{{Buras}, {Janka}, {Rampp}  \&
  {Kifonidis}}{{Buras} et~al.}{2006b}]{buras_06b}
{Buras} R.,  {Janka} H.-T.,  {Rampp} M.,   {Kifonidis} K.,  2006b, \mn@doi
  [\aap] {10.1051/0004-6361:20054654}, \href
  {http://adsabs.harvard.edu/abs/2006A%26A...457..281B} {457, 281}

\bibitem[\protect\citeauthoryear{{Burrows}}{{Burrows}}{2013}]{2013RvMP...85..245B}
{Burrows} A.,  2013, \mn@doi [Reviews of Modern Physics]
  {10.1103/RevModPhys.85.245}, \href
  {http://adsabs.harvard.edu/abs/2013RvMP...85..245B} {85, 245}

\bibitem[\protect\citeauthoryear{{Burrows}, {Hayes}  \& {Fryxell}}{{Burrows}
  et~al.}{1995}]{burrows_95}
{Burrows} A.,  {Hayes} J.,   {Fryxell} B.~A.,  1995, \mn@doi [\apj]
  {10.1086/176188}, \href {http://adsabs.harvard.edu/abs/1995ApJ...450..830B}
  {450, 830}

\bibitem[\protect\citeauthoryear{{Cerd{\'a}-Dur{\'a}n}, {DeBrye}, {Aloy},
  {Font}  \& {Obergaulinger}}{{Cerd{\'a}-Dur{\'a}n}
  et~al.}{2013}]{2013ApJ...779L..18C}
{Cerd{\'a}-Dur{\'a}n} P.,  {DeBrye} N.,  {Aloy} M.~A.,  {Font} J.~A.,
  {Obergaulinger} M.,  2013, \mn@doi [\apjl] {10.1088/2041-8205/779/2/L18},
  \href {http://adsabs.harvard.edu/abs/2013ApJ...779L..18C} {779, L18}

\bibitem[\protect\citeauthoryear{{Chini}, {Hoffmeister}, {Nasseri}, {Stahl}  \&
  {Zinnecker}}{{Chini} et~al.}{2012}]{2012MNRAS.424.1925C}
{Chini} R.,  {Hoffmeister} V.~H.,  {Nasseri} A.,  {Stahl} O.,   {Zinnecker} H.,
   2012, \mn@doi [\mnras] {10.1111/j.1365-2966.2012.21317.x}, \href
  {http://adsabs.harvard.edu/abs/2012MNRAS.424.1925C} {424, 1925}

\bibitem[\protect\citeauthoryear{{Dessart}, {Burrows}, {Livne}  \&
  {Ott}}{{Dessart} et~al.}{2006}]{dessart_06}
{Dessart} L.,  {Burrows} A.,  {Livne} E.,   {Ott} C.~D.,  2006, \mn@doi [\apj]
  {10.1086/504068}, \href {http://adsabs.harvard.edu/abs/2006ApJ...645..534D}
  {645, 534}

\bibitem[\protect\citeauthoryear{{Dimmelmeier}, {Font}  \&
  {M{\"u}ller}}{{Dimmelmeier} et~al.}{2002}]{2002A&A...393..523D}
{Dimmelmeier} H.,  {Font} J.~A.,   {M{\"u}ller} E.,  2002, \mn@doi [\aap]
  {10.1051/0004-6361:20021053}, \href
  {http://adsabs.harvard.edu/abs/2002A%26A...393..523D} {393, 523}

\bibitem[\protect\citeauthoryear{{Dimmelmeier}, {Ott}, {Marek}  \&
  {Janka}}{{Dimmelmeier} et~al.}{2008}]{2008PhRvD..78f4056D}
{Dimmelmeier} H.,  {Ott} C.~D.,  {Marek} A.,   {Janka} H.-T.,  2008, \mn@doi
  [\prd] {10.1103/PhysRevD.78.064056}, \href
  {http://esoads.eso.org/abs/2008PhRvD..78f4056D} {78, 064056}

\bibitem[\protect\citeauthoryear{{Fischer}, {Mart{\'{\i}}nez-Pinedo}, {Wu},
  {Lohs}  \& {Qian}}{{Fischer} et~al.}{2018}]{fischer_18}
{Fischer} T.,  {Mart{\'{\i}}nez-Pinedo} G.,  {Wu} M.-R.,  {Lohs} A.,   {Qian}
  Y.-Z.,  2018, preprint, \href
  {http://adsabs.harvard.edu/abs/2018arXiv180410890F} {} (\mn@eprint {arXiv}
  {1804.10890})

\bibitem[\protect\citeauthoryear{{Foglizzo}, {Galletti}, {Scheck}  \&
  {Janka}}{{Foglizzo} et~al.}{2007}]{2007ApJ...654.1006F}
{Foglizzo} T.,  {Galletti} P.,  {Scheck} L.,   {Janka} H.-T.,  2007, \mn@doi
  [\apj] {10.1086/509612}, \href
  {http://adsabs.harvard.edu/abs/2007ApJ...654.1006F} {654, 1006}

\bibitem[\protect\citeauthoryear{{Fuller}, {Klion}, {Abdikamalov}  \&
  {Ott}}{{Fuller} et~al.}{2015}]{fuller_15}
{Fuller} J.,  {Klion} H.,  {Abdikamalov} E.,   {Ott} C.~D.,  2015, \mn@doi
  [\mnras] {10.1093/mnras/stv698}, \href
  {http://adsabs.harvard.edu/abs/2015MNRAS.450..414F} {450, 414}

\bibitem[\protect\citeauthoryear{{Gill}, {Wang}, {Valdez}, {Szczepanczyk},
  {Zanolin}  \& {Mukherjee}}{{Gill} et~al.}{2018}]{2018arXiv180207255G}
{Gill} K.,  {Wang} W.,  {Valdez} O.,  {Szczepanczyk} M.,  {Zanolin} M.,
  {Mukherjee} S.,  2018, preprint, \href
  {http://adsabs.harvard.edu/abs/2018arXiv180207255G} {} (\mn@eprint {arXiv}
  {1802.07255})

\bibitem[\protect\citeauthoryear{{Glas}, {Janka}, {Melson}, {Stockinger}  \&
  {Just}}{{Glas} et~al.}{2018}]{glas_18}
{Glas} R.,  {Janka} H.-T.,  {Melson} T.,  {Stockinger} G.,   {Just} O.,  2018,
  preprint, \href {http://adsabs.harvard.edu/abs/2018arXiv180910150G} {}
  (\mn@eprint {arXiv} {1809.10150})

\bibitem[\protect\citeauthoryear{{Goldreich} \& {Kumar}}{{Goldreich} \&
  {Kumar}}{1990}]{goldreich_90}
{Goldreich} P.,  {Kumar} P.,  1990, \mn@doi [\apj] {10.1086/169376}, \href
  {http://adsabs.harvard.edu/abs/1990ApJ...363..694G} {363, 694}

\bibitem[\protect\citeauthoryear{{Gossan}, {Sutton}, {Stuver}, {Zanolin},
  {Gill}  \& {Ott}}{{Gossan} et~al.}{2016}]{2016PhRvD..93d2002G}
{Gossan} S.~E.,  {Sutton} P.,  {Stuver} A.,  {Zanolin} M.,  {Gill} K.,   {Ott}
  C.~D.,  2016, \mn@doi [\prd] {10.1103/PhysRevD.93.042002}, \href
  {http://adsabs.harvard.edu/abs/2016PhRvD..93d2002G} {93, 042002}

\bibitem[\protect\citeauthoryear{{Herant}, {Benz}, {Hix}, {Fryer}  \&
  {Colgate}}{{Herant} et~al.}{1994}]{herant_94}
{Herant} M.,  {Benz} W.,  {Hix} W.~R.,  {Fryer} C.~L.,   {Colgate} S.~A.,
  1994, \mn@doi [\apj] {10.1086/174817}, \href
  {http://adsabs.harvard.edu/abs/1994ApJ...435..339H} {435, 339}

\bibitem[\protect\citeauthoryear{{Hild}, {Chelkowski}  \& {Freise}}{{Hild}
  et~al.}{2008}]{2008arXiv0810.0604H}
{Hild} S.,  {Chelkowski} S.,   {Freise} A.,  2008, preprint, \href
  {http://adsabs.harvard.edu/abs/2008arXiv0810.0604H} {} (\mn@eprint {arXiv}
  {0810.0604})

\bibitem[\protect\citeauthoryear{{Hobbs}, {Alberg}  \& {Miller}}{{Hobbs}
  et~al.}{2016}]{hobbs_16}
{Hobbs} T.~J.,  {Alberg} M.,   {Miller} G.~A.,  2016, \mn@doi [\prc]
  {10.1103/PhysRevC.93.052801}, \href
  {http://adsabs.harvard.edu/abs/2016PhRvC..93e2801H} {93, 052801}

\bibitem[\protect\citeauthoryear{{Horowitz}}{{Horowitz}}{2002}]{horowitz_02}
{Horowitz} C.~J.,  2002, \mn@doi [\prd] {10.1103/PhysRevD.65.043001}, \href
  {http://adsabs.harvard.edu/abs/2002PhRvD..65d3001H} {65, 043001}

\bibitem[\protect\citeauthoryear{{Horowitz}, {Caballero}, {Lin}, {O'Connor}  \&
  {Schwenk}}{{Horowitz} et~al.}{2017}]{horowitz_17}
{Horowitz} C.~J.,  {Caballero} O.~L.,  {Lin} Z.,  {O'Connor} E.,   {Schwenk}
  A.,  2017, \mn@doi [\prc] {10.1103/PhysRevC.95.025801}, \href
  {http://adsabs.harvard.edu/abs/2017PhRvC..95b5801H} {95, 025801}

\bibitem[\protect\citeauthoryear{{Janka}}{{Janka}}{2012}]{2012ARNPS..62..407J}
{Janka} H.-T.,  2012, \mn@doi [Annual Review of Nuclear and Particle Science]
  {10.1146/annurev-nucl-102711-094901}, \href
  {http://adsabs.harvard.edu/abs/2012ARNPS..62..407J} {62, 407}

\bibitem[\protect\citeauthoryear{{Janka} \& {M\"uller}}{{Janka} \&
  {M\"uller}}{1995}]{janka_95}
{Janka} H.-T.,  {M\"uller} E.,  1995, \mn@doi [\apjl] {10.1086/309604}, \href
  {http://adsabs.harvard.edu/abs/1995ApJ...448L.109J} {448, L109}

\bibitem[\protect\citeauthoryear{{Just}, {Bollig}, {Janka}, {Obergaulinger},
  {Glas}  \& {Nagataki}}{{Just} et~al.}{2018}]{just_18}
{Just} O.,  {Bollig} R.,  {Janka} H.-T.,  {Obergaulinger} M.,  {Glas} R.,
  {Nagataki} S.,  2018, \mn@doi [\mnras] {10.1093/mnras/sty2578}, \href
  {http://adsabs.harvard.edu/abs/2018MNRAS.481.4786J} {481, 4786}

\bibitem[\protect\citeauthoryear{{Keil}, {Janka}  \& {Mueller}}{{Keil}
  et~al.}{1996}]{keil_96}
{Keil} W.,  {Janka} H.-T.,   {Mueller} E.,  1996, \mn@doi [\apjl]
  {10.1086/310404}, \href {http://adsabs.harvard.edu/abs/1996ApJ...473L.111K}
  {473, L111}

\bibitem[\protect\citeauthoryear{{Kotake}}{{Kotake}}{2013}]{kotake_13}
{Kotake} K.,  2013, \mn@doi [Comptes Rendus Physique]
  {10.1016/j.crhy.2013.01.008}, \href
  {http://adsabs.harvard.edu/abs/2013CRPhy..14..318K} {14, 318}

\bibitem[\protect\citeauthoryear{{Kotake} \& {Kuroda}}{{Kotake} \&
  {Kuroda}}{2017}]{kotake_handbook}
{Kotake} K.,  {Kuroda} T.,  2017, in {Alsabti} A.~W.,  {Murdin} P.,  eds, ,
  Handbook of Supernovae.
pringer International Publishing AG, p.~1671,
  \mn@doi{10.1007/978-3-319-21846-5_9}

\bibitem[\protect\citeauthoryear{Kumar, Chatterjee  \& Verma}{Kumar
  et~al.}{2014}]{kumar_14}
Kumar A.,  Chatterjee A.~G.,   Verma M.~K.,  2014, \mn@doi [Phys. Rev. E]
  {10.1103/PhysRevE.90.023016}, 90, 023016

\bibitem[\protect\citeauthoryear{{Kuroda}, {Kotake}  \& {Takiwaki}}{{Kuroda}
  et~al.}{2016}]{2016ApJ...829L..14K}
{Kuroda} T.,  {Kotake} K.,   {Takiwaki} T.,  2016, \mn@doi [\apjl]
  {10.3847/2041-8205/829/1/L14}, \href
  {http://adsabs.harvard.edu/abs/2016ApJ...829L..14K} {829, L14}

\bibitem[\protect\citeauthoryear{{Lattimer} \& {Mazurek}}{{Lattimer} \&
  {Mazurek}}{1981}]{lattimer_81}
{Lattimer} J.~M.,  {Mazurek} T.~J.,  1981, \mn@doi [\apj] {10.1086/158989},
  \href {http://adsabs.harvard.edu/abs/1981ApJ...246..955L} {246, 955}

\bibitem[\protect\citeauthoryear{Lattimer \& Swesty}{Lattimer \&
  Swesty}{1991}]{Lattimer:1991nc}
Lattimer J.~M.,  Swesty F.~D.,  1991, \mn@doi [Nucl. Phys.]
  {10.1016/0375-9474(91)90452-C}, A535, 331

\bibitem[\protect\citeauthoryear{Lynch, Vitale, Essick, Katsavounidis  \&
  Robinet}{Lynch et~al.}{2017}]{PhysRevD.95.104046}
Lynch R.,  Vitale S.,  Essick R.,  Katsavounidis E.,   Robinet F.,  2017,
  \mn@doi [Phys. Rev. D] {10.1103/PhysRevD.95.104046}, 95, 104046

\bibitem[\protect\citeauthoryear{{Marek}}{{Marek}}{2007}]{marek_phd}
{Marek} A.,  2007, {PhD thesis}, Technische Universti\"at M\"unchen

\bibitem[\protect\citeauthoryear{{Mart{\'{\i}}nez-Pinedo}, {Fischer}, {Lohs}
  \& {Huther}}{{Mart{\'{\i}}nez-Pinedo} et~al.}{2012}]{martinez_12}
{Mart{\'{\i}}nez-Pinedo} G.,  {Fischer} T.,  {Lohs} A.,   {Huther} L.,  2012,
  \mn@doi [Physical Review Letters] {10.1103/PhysRevLett.109.251104}, \href
  {http://adsabs.harvard.edu/abs/2012PhRvL.109y1104M} {109, 251104}

\bibitem[\protect\citeauthoryear{{Melson}, {Janka}, {Bollig}, {Hanke}, {Marek}
  \& {M{\"u}ller}}{{Melson} et~al.}{2015}]{melson_15}
{Melson} T.,  {Janka} H.-T.,  {Bollig} R.,  {Hanke} F.,  {Marek} A.,
  {M{\"u}ller} B.,  2015, \mn@doi [\apjl] {10.1088/2041-8205/808/2/L42}, \href
  {http://adsabs.harvard.edu/abs/2015ApJ...808L..42M} {808, L42}

\bibitem[\protect\citeauthoryear{{Mirizzi}, {Tamborra}, {Janka}, {Saviano},
  {Scholberg}, {Bollig}, {H{\"u}depohl}  \& {Chakraborty}}{{Mirizzi}
  et~al.}{2016}]{mirizzi_16}
{Mirizzi} A.,  {Tamborra} I.,  {Janka} H.-T.,  {Saviano} N.,  {Scholberg} K.,
  {Bollig} R.,  {H{\"u}depohl} L.,   {Chakraborty} S.,  2016, \mn@doi [Nuovo
  Cimento Rivista Serie] {10.1393/ncr/i2016-10120-8}, \href
  {http://adsabs.harvard.edu/abs/2016NCimR..39....1M} {39, 1}

\bibitem[\protect\citeauthoryear{{Morozova}, {Radice}, {Burrows}  \&
  {Vartanyan}}{{Morozova} et~al.}{2018}]{morozova_18}
{Morozova} V.,  {Radice} D.,  {Burrows} A.,   {Vartanyan} D.,  2018, \mn@doi
  [\apj] {10.3847/1538-4357/aac5f1}, \href
  {http://adsabs.harvard.edu/abs/2018ApJ...861...10M} {861, 10}

\bibitem[\protect\citeauthoryear{{Mosser} et~al.,}{{Mosser}
  et~al.}{2012}]{2012A&A...548A..10M}
{Mosser} B.,  et~al., 2012, \mn@doi [\aap] {10.1051/0004-6361/201220106}, \href
  {http://cdsads.u-strasbg.fr/abs/2012A%26A...548A..10M} {548, A10}

\bibitem[\protect\citeauthoryear{{Mueller} \& {Janka}}{{Mueller} \&
  {Janka}}{1997}]{1997A&A...317..140M}
{Mueller} E.,  {Janka} H.-T.,  1997, \aap, \href
  {http://adsabs.harvard.edu/abs/1997A%26A...317..140M} {317, 140}

\bibitem[\protect\citeauthoryear{{M{\"u}ller}}{{M{\"u}ller}}{2015}]{mueller_15b}
{M{\"u}ller} B.,  2015, \mn@doi [\mnras] {10.1093/mnras/stv1611}, \href
  {http://adsabs.harvard.edu/abs/2015MNRAS.453..287M} {453, 287}

\bibitem[\protect\citeauthoryear{{M{\"u}ller}}{{M{\"u}ller}}{2017}]{mueller_17}
{M{\"u}ller} B.,  2017, arXiv e-prints, \href
  {http://adsabs.harvard.edu/abs/2017arXiv170304633M} {}

\bibitem[\protect\citeauthoryear{{M{\"u}ller} \& {Janka}}{{M{\"u}ller} \&
  {Janka}}{2014}]{mueller_14}
{M{\"u}ller} B.,  {Janka} H.-T.,  2014, \mn@doi [\apj]
  {10.1088/0004-637X/788/1/82}, \href
  {http://adsabs.harvard.edu/abs/2014ApJ...788...82M} {788, 82}

\bibitem[\protect\citeauthoryear{{M{\"u}ller} \& {Janka}}{{M{\"u}ller} \&
  {Janka}}{2015}]{mueller_15a}
{M{\"u}ller} B.,  {Janka} H.-T.,  2015, \mn@doi [\mnras]
  {10.1093/mnras/stv101}, \href
  {http://adsabs.harvard.edu/abs/2015MNRAS.448.2141M} {448, 2141}

\bibitem[\protect\citeauthoryear{{M{\"u}ller}, {Janka}  \&
  {Dimmelmeier}}{{M{\"u}ller} et~al.}{2010}]{2010ApJS..189..104M}
{M{\"u}ller} B.,  {Janka} H.-T.,   {Dimmelmeier} H.,  2010, \mn@doi [\apjs]
  {10.1088/0067-0049/189/1/104}, \href
  {http://adsabs.harvard.edu/abs/2010ApJS..189..104M} {189, 104}

\bibitem[\protect\citeauthoryear{{M{\"u}ller}, {Janka}  \&
  {Wongwathanarat}}{{M{\"u}ller} et~al.}{2012}]{2012A&A...537A..63M}
{M{\"u}ller} E.,  {Janka} H.-T.,   {Wongwathanarat} A.,  2012, \mn@doi [\aap]
  {10.1051/0004-6361/201117611}, \href
  {http://esoads.eso.org/abs/2012A%26A...537A..63M} {537, A63}

\bibitem[\protect\citeauthoryear{{M{\"u}ller}, {Janka}  \&
  {Marek}}{{M{\"u}ller} et~al.}{2013}]{2013ApJ...766...43M}
{M{\"u}ller} B.,  {Janka} H.-T.,   {Marek} A.,  2013, \mn@doi [\apj]
  {10.1088/0004-637X/766/1/43}, \href
  {http://adsabs.harvard.edu/abs/2013ApJ...766...43M} {766, 43}

\bibitem[\protect\citeauthoryear{{M{\"u}ller}, {Viallet}, {Heger}  \&
  {Janka}}{{M{\"u}ller} et~al.}{2016}]{2016ApJ...833..124M}
{M{\"u}ller} B.,  {Viallet} M.,  {Heger} A.,   {Janka} H.-T.,  2016, \mn@doi
  [\apj] {10.3847/1538-4357/833/1/124}, \href
  {http://adsabs.harvard.edu/abs/2016ApJ...833..124M} {833, 124}

\bibitem[\protect\citeauthoryear{{M{\"u}ller}, {Melson}, {Heger}  \&
  {Janka}}{{M{\"u}ller} et~al.}{2017}]{2017MNRAS.472..491M}
{M{\"u}ller} B.,  {Melson} T.,  {Heger} A.,   {Janka} H.-T.,  2017, \mn@doi
  [\mnras] {10.1093/mnras/stx1962}, \href
  {http://adsabs.harvard.edu/abs/2017MNRAS.472..491M} {472, 491}

\bibitem[\protect\citeauthoryear{{M{\"u}ller} et~al.,}{{M{\"u}ller}
  et~al.}{2019}]{2018arXiv181105483M}
{M{\"u}ller} B.,  et~al., 2019, \mn@doi [\mnras] {10.1093/mnras/stz216}, \href
  {https://ui.adsabs.harvard.edu/abs/2019MNRAS.484.3307M} {484, 3307}

\bibitem[\protect\citeauthoryear{{Murphy}, {Ott}  \& {Burrows}}{{Murphy}
  et~al.}{2009}]{2009ApJ...707.1173M}
{Murphy} J.~W.,  {Ott} C.~D.,   {Burrows} A.,  2009, \mn@doi [\apj]
  {10.1088/0004-637X/707/2/1173}, \href
  {http://adsabs.harvard.edu/abs/2009ApJ...707.1173M} {707, 1173}

\bibitem[\protect\citeauthoryear{{Murphy}, {Dolence}  \& {Burrows}}{{Murphy}
  et~al.}{2013}]{murphy_12}
{Murphy} J.~W.,  {Dolence} J.~C.,   {Burrows} A.,  2013, \mn@doi [\apj]
  {10.1088/0004-637X/771/1/52}, \href
  {http://adsabs.harvard.edu/abs/2013ApJ...771...52M} {771, 52}

\bibitem[\protect\citeauthoryear{{O'Connor} \& {Couch}}{{O'Connor} \&
  {Couch}}{2018}]{oconnor_18}
{O'Connor} E.~P.,  {Couch} S.~M.,  2018, \mn@doi [\apj]
  {10.3847/1538-4357/aadcf7}, \href
  {http://adsabs.harvard.edu/abs/2018ApJ...865...81O} {865, 81}

\bibitem[\protect\citeauthoryear{{Obukhov}}{{Obukhov}}{1959}]{obukhov_59}
{Obukhov} A.~M.,  1959, \mn@doi [Dokl.~Akad.~Nauk SSSR]
  {10.1016/S0065-2687(08)60098-9}, 125, 124

\bibitem[\protect\citeauthoryear{{Ott} et~al.,}{{Ott}
  et~al.}{2013}]{2013ApJ...768..115O}
{Ott} C.~D.,  et~al., 2013, \mn@doi [\apj] {10.1088/0004-637X/768/2/115}, \href
  {https://ui.adsabs.harvard.edu/#abs/2013ApJ...768..115O} {768, 115}

\bibitem[\protect\citeauthoryear{{Powell}, {Gossan}, {Logue}  \&
  {Heng}}{{Powell} et~al.}{2016}]{2016PhRvD..94l3012P}
{Powell} J.,  {Gossan} S.~E.,  {Logue} J.,   {Heng} I.~S.,  2016, \mn@doi
  [\prd] {10.1103/PhysRevD.94.123012}, \href
  {http://adsabs.harvard.edu/abs/2016PhRvD..94l3012P} {94, 123012}

\bibitem[\protect\citeauthoryear{{Powell}, {Szczepanczyk}  \& {Heng}}{{Powell}
  et~al.}{2017}]{2017PhRvD..96l3013P}
{Powell} J.,  {Szczepanczyk} M.,   {Heng} I.~S.,  2017, \mn@doi [\prd]
  {10.1103/PhysRevD.96.123013}, \href
  {http://adsabs.harvard.edu/abs/2017PhRvD..96l3013P} {96, 123013}

\bibitem[\protect\citeauthoryear{Punturo et~al.,}{Punturo
  et~al.}{2010}]{0264-9381-27-19-194002}
Punturo M.,  et~al., 2010, Classical and Quantum Gravity, 27, 194002

\bibitem[\protect\citeauthoryear{{Radice}, {Morozova}, {Burrows}, {Vartanyan}
  \& {Nagakura}}{{Radice} et~al.}{2018}]{radice_18}
{Radice} D.,  {Morozova} V.,  {Burrows} A.,  {Vartanyan} D.,   {Nagakura} H.,
  2018, arXiv e-prints, \href
  {http://adsabs.harvard.edu/abs/2018arXiv181207703R} {}

\bibitem[\protect\citeauthoryear{{Rampp} \& {Janka}}{{Rampp} \&
  {Janka}}{2002}]{rampp_02}
{Rampp} M.,  {Janka} H.-T.,  2002, \mn@doi [\aap] {10.1051/0004-6361:20021398},
  \href {http://adsabs.harvard.edu/abs/2002A%26A...396..361R} {396, 361}

\bibitem[\protect\citeauthoryear{{Richers}, {Ott}, {Abdikamalov}, {O'Connor}
  \& {Sullivan}}{{Richers} et~al.}{2017}]{2017PhRvD..95f3019R}
{Richers} S.,  {Ott} C.~D.,  {Abdikamalov} E.,  {O'Connor} E.,   {Sullivan} C.,
   2017, \mn@doi [\prd] {10.1103/PhysRevD.95.063019}, \href
  {http://adsabs.harvard.edu/abs/2017PhRvD..95f3019R} {95, 063019}

\bibitem[\protect\citeauthoryear{{Sana} et~al.,}{{Sana}
  et~al.}{2013}]{2013ASPC..470..141S}
{Sana} H.,  et~al., 2013, in {Pugliese} G.,  {de Koter} A.,   {Wijburg} M.,
  eds,  Astronomical Society of the Pacific Conference Series Vol. 470, 370
  Years of Astronomy in Utrecht. p.~141 (\mn@eprint {arXiv} {1211.4740})

\bibitem[\protect\citeauthoryear{{Scheidegger}, {K{\"a}ppeli}, {Whitehouse},
  {Fischer}  \& {Liebend{\"o}rfer}}{{Scheidegger}
  et~al.}{2010}]{2010A&A...514A..51S}
{Scheidegger} S.,  {K{\"a}ppeli} R.,  {Whitehouse} S.~C.,  {Fischer} T.,
  {Liebend{\"o}rfer} M.,  2010, \mn@doi [\aap] {10.1051/0004-6361/200913220},
  \href {http://adsabs.harvard.edu/abs/2010A%26A...514A..51S} {514, A51}

\bibitem[\protect\citeauthoryear{{Smith} \& {Thrane}}{{Smith} \&
  {Thrane}}{2018}]{2018PhRvX...8b1019S}
{Smith} R.,  {Thrane} E.,  2018, \mn@doi [Physical Review X]
  {10.1103/PhysRevX.8.021019}, \href
  {http://adsabs.harvard.edu/abs/2018PhRvX...8b1019S} {8, 021019}

\bibitem[\protect\citeauthoryear{{Takiwaki} \& {Kotake}}{{Takiwaki} \&
  {Kotake}}{2018}]{2018MNRAS.475L..91T}
{Takiwaki} T.,  {Kotake} K.,  2018, \mn@doi [\mnras] {10.1093/mnrasl/sly008},
  \href {http://adsabs.harvard.edu/abs/2018MNRAS.475L..91T} {475, L91}

\bibitem[\protect\citeauthoryear{{Tamborra}, {Hanke}, {Janka}, {M{\"u}ller},
  {Raffelt}  \& {Marek}}{{Tamborra} et~al.}{2014}]{tamborra_14}
{Tamborra} I.,  {Hanke} F.,  {Janka} H.-T.,  {M{\"u}ller} B.,  {Raffelt} G.~G.,
    {Marek} A.,  2014, \mn@doi [\apj] {10.1088/0004-637X/792/2/96}, \href
  {http://adsabs.harvard.edu/abs/2014ApJ...792...96T} {792, 96}

\bibitem[\protect\citeauthoryear{{Tauris}, {Langer}, {Moriya}, {Podsiadlowski},
  {Yoon}  \& {Blinnikov}}{{Tauris} et~al.}{2013}]{2013ApJ...778L..23T}
{Tauris} T.~M.,  {Langer} N.,  {Moriya} T.~J.,  {Podsiadlowski} P.,  {Yoon}
  S.~C.,   {Blinnikov} S.~I.,  2013, \mn@doi [\apj]
  {10.1088/2041-8205/778/2/L23}, \href
  {https://ui.adsabs.harvard.edu/#abs/2013ApJ...778L..23T} {778, L23}

\bibitem[\protect\citeauthoryear{{Tauris}, {Langer}  \&
  {Podsiadlowski}}{{Tauris} et~al.}{2015}]{2015MNRAS.451.2123T}
{Tauris} T.~M.,  {Langer} N.,   {Podsiadlowski} P.,  2015, \mn@doi [\mnras]
  {10.1093/mnras/stv990}, \href
  {http://adsabs.harvard.edu/abs/2015MNRAS.451.2123T} {451, 2123}

\bibitem[\protect\citeauthoryear{{The LIGO Scientific Collaboration}, {Aasi},
  {Abbott}, {Abbott}  \& et al.}{{The LIGO Scientific Collaboration}
  et~al.}{2015}]{aLIGO}
{The LIGO Scientific Collaboration} {Aasi} J.,  {Abbott} B.~P.,  {Abbott} R.,
  et al. 2015, \mn@doi [Classical and Quantum Gravity]
  {10.1088/0264-9381/32/7/074001}, \href
  {http://adsabs.harvard.edu/abs/2015CQGra..32g4001T} {32, 074001}

\bibitem[\protect\citeauthoryear{{Torres-Forn{\'e}}, {Cerd{\'a}-Dur{\'a}n},
  {Passamonti}  \& {Font}}{{Torres-Forn{\'e}}
  et~al.}{2018}]{2018MNRAS.474.5272T}
{Torres-Forn{\'e}} A.,  {Cerd{\'a}-Dur{\'a}n} P.,  {Passamonti} A.,   {Font}
  J.~A.,  2018, \mn@doi [\mnras] {10.1093/mnras/stx3067}, \href
  {http://adsabs.harvard.edu/abs/2018MNRAS.474.5272T} {474, 5272}

\bibitem[\protect\citeauthoryear{{Wellstein}, {Langer}  \& {Braun}}{{Wellstein}
  et~al.}{2001}]{2001A&A...369..939W}
{Wellstein} S.,  {Langer} N.,   {Braun} H.,  2001, \mn@doi [\aap]
  {10.1051/0004-6361:20010151}, \href
  {http://adsabs.harvard.edu/abs/2001A%26A...369..939W} {369, 939}

\bibitem[\protect\citeauthoryear{{Yakunin} et~al.,}{{Yakunin}
  et~al.}{2010}]{yakunin_10}
{Yakunin} K.~N.,  et~al., 2010, \mn@doi [Classical and Quantum Gravity]
  {10.1088/0264-9381/27/19/194005}, \href
  {http://adsabs.harvard.edu/abs/2010CQGra..27s4005Y} {27, 194005}

\bibitem[\protect\citeauthoryear{{Yakunin} et~al.,}{{Yakunin}
  et~al.}{2015}]{yakunin_16}
{Yakunin} K.~N.,  et~al., 2015, \mn@doi [\prd] {10.1103/PhysRevD.92.084040},
  \href {http://adsabs.harvard.edu/abs/2015PhRvD..92h4040Y} {92, 084040}

\bibitem[\protect\citeauthoryear{{Yakunin} et~al.,}{{Yakunin}
  et~al.}{2017}]{2017arXiv170107325Y}
{Yakunin} K.~N.,  et~al., 2017, preprint, \href
  {http://adsabs.harvard.edu/abs/2017arXiv170107325Y} {} (\mn@eprint {arXiv}
  {1701.07325})

\makeatother
\end{thebibliography}


\appendix


\bsp	
\label{lastpage}
\end{document}